\documentclass[useAMS,usegraphicx,usenatbib]{mn2e}

\usepackage{multirow}
\bibpunct{(}{)}{;}{a}{}{,}

\title[$\alpha_{\rm OX}$ measurements in GBHs]{What can we learn about
  Quasars from $\alpha_{\rm OX}$ measurements in Galactic Black Hole
  Binaries?} 

\author[Sobolewska et
  al.]{Ma{\l}gorzata A.
  Sobolewska$^{1, 2}$\thanks{E-mail:malgosia@physics.uoc.gr},
  Marek Gierli\'nski$^1$, Aneta Siemiginowska$^3$\\$^1$Department of
  Physics, University of Durham, South Road, Durham DH1 3LE,
  UK\\$^2$Foundation for Research and Technology--Hellas/IESL,
  Voutes, 71110 Heraklion, Crete, Greece\\$^3$Smithsonian Astrophysical Observatory, 60 Garden Street,
  Cambridge, MA 02138, USA}
\newcommand{\aj}{AJ}
\newcommand{\apj}{ApJ}
\newcommand{\apjl}{ApJL}
\newcommand{\mnras}{MNRAS}
\newcommand{\aap}{A\&A}

\newcommand{\pasj}{PASJ}
\newcommand{\araa}{ARAA}
\newcommand{\apjs}{ApJS}
\newcommand{\aspc}{ASP Conf. Ser. Vol. }
\newcommand{\asp}{Astron. Soc. Pac., San Francisco}
\newcommand{\aapr}{A\&AR}

\begin{document}

\bibliographystyle{plainnat}
\date{}

\pagerange{\pageref{firstpage}--\pageref{lastpage}} \pubyear{2008}

\maketitle

\label{firstpage}

\begin{abstract}
We draw a comparison between AGN and Galactic
black hole binaries using a uniform description of spectral energy
distribution of these two classes of X-ray sources. We parametrize
spectra of GBHs with an $\alpha_{\rm GBH}$ parameter which we define as a slope of
a nominal power law function between 3 and 20 keV. We show that this
parameter can be treated as an equivalent of the X-ray loudness,
$\alpha_{\rm OX}$, used to describe AGN spectra. We do not find linear
correlation between the $\alpha_{\rm GBH}$ and disc flux (similar to that
between $\alpha_{\rm OX}$ and optical/UV luminosity found in AGN). Instead,
we show that $\alpha_{\rm GBH}$ follows a well defined pattern during a
GBH outburst. We find that $\alpha_{\rm GBH}$ tend to cluster around 1, 1.5
and 2, which correspond to a hard, very
high/intermediate and soft spectral state, respectively. We
conclude that majority of the observed Type 1 radio quiet AGN are in the
spectral state corresponding to the very high/intermediate state of
GBHs. The same conclusion is valid for radio loud AGN. We also study variations of the spectral slopes ($\alpha_{\rm GBH}$
and the X-ray photon index, $\Gamma$) as a function of disc and
Comptonization fluxes. We discuss these dependencies in the context of
correlations of $\alpha_{\rm OX}$ and $\Gamma$ with the optical/UV and
X-ray 2 keV fluxes considered for AGN and quasars.
\end{abstract}



\section{Introduction}
\label{sec:intro}

Accretion powers active galactic nuclei (AGN), including quasars, and
drives X-ray activity of Galactic black hole binaries (GBHs). While GBHs
harbour a few solar mass black hole, a super-massive black hole
($>10^{6-10}$ M$_{\odot}$) is needed to power quasars. Spectral energy
distribution (SED) of AGN in optical to X-ray band has two main
components. A multi-colour disc blackbody is believed to originate in the
optically thick matter creating an accretion disc around the black hole
\citep[e.g.][]{mitsudaea:1984}, 
and hard X-ray power-law like radiation is thought to come from an
optically thin hot electron plasma, so-called 'corona'. The location and
origins of the 'corona' is not clear. A number of theoretical models were
invoked to explain this hard X-ray radiation including inverse Compton
process in a 'corona' above the disc \citep[e.g.][]{coppi:1999} or jet-related
synchrotron radiation \citep*[e.g.][]{markoffea:2005}. There is 
an ongoing debate as to which model describes the hard X-rays better
\citep[e.g.][]{zdziarskiea:2004,markoffea:2005}. 

Characteristic disc temperature scales with mass as $M^{-1/4}$
for a given fraction of Eddington luminosity, so the disc is hotter in 
GBHs and emits in soft X-ray band instead of optical/UV band. Hence, all
levels of GBHs activity can be studied in the X-ray band.  {\it 
Rossi X-ray Timing Explorer} ({\it RXTE\/}) monitoring campaigns
provided with excellent X-ray data from GBHs. High Energy X-ray Timing
Experiment (HEXTE) on board {\it RXTE} can observe bright sources up
to $\sim$ 200 keV.  {\it BeppoSAX}, {\it INTEGRAL} and {\it Suzaku}
provide with excellent broad-band spectral coverage. In contrast to
these data, distant quasars have fluxes too low to be detected by {\it
RXTE}. Their accretion disc radiation is shifted into the optical/UV
band and needs to be observed with optical and UV instruments. In
X-rays, the most recent {\it Chandra} and {\it XMM-Newton}
observatories provide X-ray continua with excellent resolution, but
only in a limited 0.3--8 keV band (1.5--40 keV in the rest frame of a
$z$ = 4 quasar).

The broad-band optical/UV/X-ray spectra of local AGN and high-$z$
quasars are usually parametrized by the monochromatic rest frame
optical/UV and X-ray luminosities, $L({\rm 2~keV})$ and $L({\rm 2500\AA})$
in erg s$^{-1}$ Hz$^{-1}$, respectively, and the X-ray 
photon index, $\Gamma$. In this paper we will use $l_{\rm UV}=\log(\nu
L_{\nu})_{\rm O}$ at $\lambda = c/\nu_{\rm O}$ = 2500~\AA\ and $l_{\rm
  X}=\log(\nu L_{\nu})_{\rm X}$ at $E = h/\nu_{\rm X}$ = 2 keV. 

The X-ray loudness is another parameter in AGN studies that
characterizes the relative amount of energy emitted in the optical
(e.g. thermal disc component) and X-rays (non-thermal emission). It is
defined as a point-to-point spectral slope between the optical/UV and
X-ray bands \citep{avni:82}:
\begin{equation}
\label{eq:eq1}
\alpha_{\rm OX} = -\frac{\log (\nu L_{\nu})_{\rm O} -\log (\nu
  L_{\nu})_{\rm X}}{\log \nu_{\rm O}-\log \nu_{\rm X}}+1 , 
\end{equation}
Equation~\ref{eq:eq1} is equivalent to that used in other papers,
$\alpha_{\rm OX} = -0.3838\log[L({\rm 2~keV})/L({\rm 2500\AA})]$. We use
the minus sign, so our $\alpha_{\rm OX}$ will be positive. 


\begin{table*}
\centering
\caption{List of sources, available RXTE data, properties of the outbursts
  and energy spectra. (1) Source name, (2) Id of an outburst if more than
  one outburst was analysed, (3--5) Year, start day and end day of an
  outburst (6) Total number of pointings analysed, and number of pointings
  that went through the criteria of Methods~1 and 2, (7) Spectral states
  (HS -- hard state, IS -- intermediate state and SS -- soft state)
  observed in Method 2 pointings, (8) Galactic absorption used in the fits
  found in literature. The numbered references (9) are as follows: [1]
  \protect{\citet*{gierlinski:2001}}; [2] \protect{\citet{gierlinskidone:2003}}; [3]
  \protect{\citet*{miniutti:2004}}; [4] \protect{\citet{hynes:2002}}; [5]
  \protect{\citet{capitanio:2005}}; [6] \protect{\citet{zdziarski:1998}}.} 
\begin{tabular}{l c c c c c c c c c c c}
\hline
Source & Outburst & Year & Start   & End     & \multicolumn{3}{c}{No. of pointings} & Observed        & $N_{\rm H}$ & References\\
       &          &      & MJD$^a$ & MJD$^a$ & Total & M1  & M2                     & spectral states & $\times$10$^{22}$ cm$^{-2}$ &\\
(1)    & (2)      & (3)  & (4)     & (5)     & \multicolumn{3}{c}{(6)}              & (7)             & (8) & (9)\\
\hline
GRO J1655-40  &    & 2005       & 53423 & 53685 & 503 & 409 & 344 & all         & 0.8 & [1]\\
XTE J1550-564 & o1 & 1998/1999  & 51064 & 51319 & 233 & 207 & 163 & all         & 0.65 & [2]\\
              & o2 & 2000       & 51644 & 51741 & \multirow{4}{*}{$\Biggr\}$164} & \multirow{4}{*}{161} & \multirow{4}{*}{103} & \multirow{4}{*}{all} & \multirow{4}{*}{0.65} & \multirow{4}{*}{[2]}\\
              & o3 & 2001       & 51938 & 52028 & & & & & &\\
              & o4 & 2002       & 52285 & 52339 & & & & & &\\
              & o5 & 2003       & 52726 & 52775 & & & & & &\\
XTE J1650-500 &    & 2001/2002  & 52159 & 52385 & 153 & 153 & 73  & all         & 0.78 & [3]\\
XTE J1859+226 &    & 1999/2000  & 51463 & 51749 & 127 & 127 & 100 & SS, IS & 0.3 & [4]\\
H 1743-322    & o1 & 2003/2004  & 52727 & 53055 & 217 & 205 & 157 & all & \multirow{3}{*}{$\Biggr\}$2.4} & \\
              & o2 & 2004       & 53198 & 53329 & \multirow{2}{*}{$\bigr\}$64} & \multirow{2}{*}{64} & \multirow{2}{*}{29} & SS & & [5] \\
              & o3 & 2005       & 53595 & 53668 & & & & IS & & \\
GX 339-4      & o1 & 1996--1999 & 50291 & 51442 & 90  & 89  & 45  & \multirow{3}{*}{$\Biggr\}$all} & &\\
              & o2 & 2002/2003  & 52367 & 52808 & 206 & 200 & 140 &  & 0.6 & [6]\\
              & o3 & 2004/2005  & 53044 & 53603 & 173 & 173 & 134 &  & &\\
\hline
\end{tabular}\\
$^a$ Modified Julian Date, ${\rm MJD} = {\rm JD} - 2400000.5$
\label{tab:tab1}
\end{table*}


\begin{figure}
\centering
\includegraphics[width=5.7cm,bb=18 144 565 680,clip]{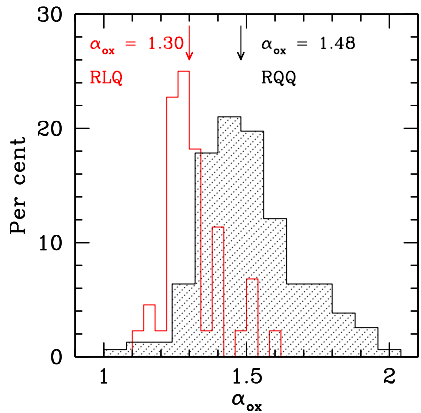}
\caption{Distribution of $\alpha_{\rm OX}$ for quasars. The filled
  histogram shows a sample of 157 optically selected AGN from
  \protect{\citet{kelly:2007}}, while the open one shows the subsample of 44
  radio-loud quasars from \protect{\citet[][]{greenea:1995}}. The medians of the two
  samples are indicated.}
\label{fig:fig1}
\end{figure}

The dependencies of $\alpha_{\rm OX}$ and $\Gamma$ on luminosity ($l_{\rm
  UV}$ and/or $l_{\rm X}$) and redshift have
important implications on our understanding of quasars nature and
evolution, and have been searched for and modeled for many years
  \citep*[e.g.][]{zamorani:81,avni:82,wilkes:1994,bechtold:2003,vignali:2003,sobol:2004a,sobol:2004b,richards:06,strateva:2005,steffen:2006,tang:2007,kelly:2007,kelly:2008}.  
 Many studies concluded that there is no evidence for a correlation between
$\alpha_{\rm OX}$ and redshift or luminosity, although some studies
reported that $\alpha_{\rm OX}$ is correlated with both.  Recently,
\cite{kelly:2008} investigated this correlation in quasars
with measured black hole mass. They found that $\alpha_{\rm OX}$
increases with increasing black hole mass and UV luminosity (in terms
of the Eddington luminosity) and decreases with increasing X-ray
luminosity (in terms of Eddington luminosity), and argue that this
implies that the fraction of bolometric luminosity emitted by an
accretion disc (thermal), as compared to a corona (non-thermal),
increases with increasing accretion rate relative to Eddington. This
may suggest that quasars and AGN should exhibit an evolution and a
variety of accretion states should be observed.  Evolution time-scales
of super-massive black holes are too long for monitoring
\citep[e.g. $>10^4$--$10^5$~years as suggested by AGN outbursts observed in
X-ray clusters, see review by][]{mcnamara:2007}, hence studies
of large samples of AGN are necessary to identify these accretion
states. 

GBHs have more favorable time-scales (days-years) for
evolution studies and many X-ray spectral states have been
observed in one source \citep[a recent review on accretion in Galactic
  black hole and neutron star binaries can be found
  in][]{doneea:2007}. These objects provide the best observational evidence  
that the accretion process is not steady even with the continuous
supply of accretion fuel to the central engine. Phenomenology of the GBHs
X-ray spectral states is well understood and so their spectra may be
used as a guidance in studies of accretion processes in AGN. A number of
methods have been developed to facilitate separation of the spectral states
in GBHs including colour-colour \citep[e.g.][]{donegier:03} and
hardness-intensity diagrams
\citep[e.g.][]{corbelea:2004,belloniea:2005,tomsickea:2005,dunnea:2008}. Generalized
version of hardness-intensity diagrams was applied to AGN by
\citet[][]{kording:06} to explore origins of the radio-loudness
(radio-to-optical flux ratio). Based on hardness-intensity diagrams
\citet*[][]{fenderea:2004} established a link between an X-ray spectral
state and radio emission from jet in X-ray binaries. In their picture, 
the low/hard state the X-ray binaries is associated with a steady compact
jet. After transition to the intermediate/very high state, the radio
emission originates in discrete ejections of plasma and it fades away while
the source continues to evolve \citep[e.g.][]{corbelea:2001}. The high/soft
spectral state is radio-quiet.

In this paper we concentrate on the X-ray loudness of AGN, optical-to-X-ray
  flux ratio, $\alpha_{\rm OX}$. Figure~\ref{fig:fig1} shows a distribution
of $\alpha_{\rm OX}$ in a sample of 157 optically selected Type 1
radio-quiet quasars \citep[RQQs,][]{kelly:2008} and in a sample of 908
radio-loud quasars observed in the soft 0.1--2.4 keV X-ray band by {\it ROSAT}
\citep[RLQs,][]{greenea:1995}. RLQs are on average more X-ray loud with
median $\alpha_{\rm  OX} = 1.30$, as opposed to median $\alpha_{\rm OX} = 1.48$
for RQQs. (Objects with only an upper limit on $\alpha_{\rm OX}$ are excluded
from the radio-quiet sample.) This difference is qualitatively explained by
  the jet radiation contributing to the X-ray band in the RLQs.

We compare $\alpha_{\rm 
  OX}$ and analogous parameter that we define for GBHs, $\alpha_{\rm
  GBH}$. The two parameters 
have the same physical meaning: they provide a measure of relative strength
of the accretion disc and Comptonized components. We study the
$\alpha_{\rm GBH}$ evolution with disc and Comptonization luminosities
  throughout an outburst to provide a baseline for comparison with
  relations studied extensively in 
AGN: $\alpha_{\rm OX}$ vs. $l_{\rm UV}$ and $l_{\rm X}$. Our goal is to use
phenomenology of GBHs to learn more about spectral states of local
AGN and high-$z$ quasars. If accretion acts in
a similar way in massive black holes in the centre of galaxies and in
stellar mass black hole binaries, the properties of $\alpha_{\rm OX}$ and
$\alpha_{\rm GBH}$ distributions, and the way they depend on other spectral
observables should correspond to each other.

The structure of the paper is as follows. In Sec.~\ref{sec:parametrization}
we define the $\alpha_{\rm GBH}$ parameter for Galactic black hole
binaries. Section~\ref{sec:data} contains details of {\it RXTE} data
selection, reduction and analysis. We describe there our results regarding 
parametrization of several outbursts of black hole binaries with $\alpha_{\rm
GBH}$. In Sec.~\ref{sec:results} we compare GBHs and AGN in terms of
spectral indices ($\alpha_{\rm GBH}$, $\alpha_{\rm OX}$ and $\Gamma$) and
their relations with the disc and Componization
fluxes. Section~\ref{sec:conclusions} contains discussion of our results 
and conclusions. 


\section{Parametrization of GBH binaries}
\label{sec:parametrization}

\begin{figure*}
\centering
\includegraphics[height=4.8cm,bb=230 500 425 675,clip]{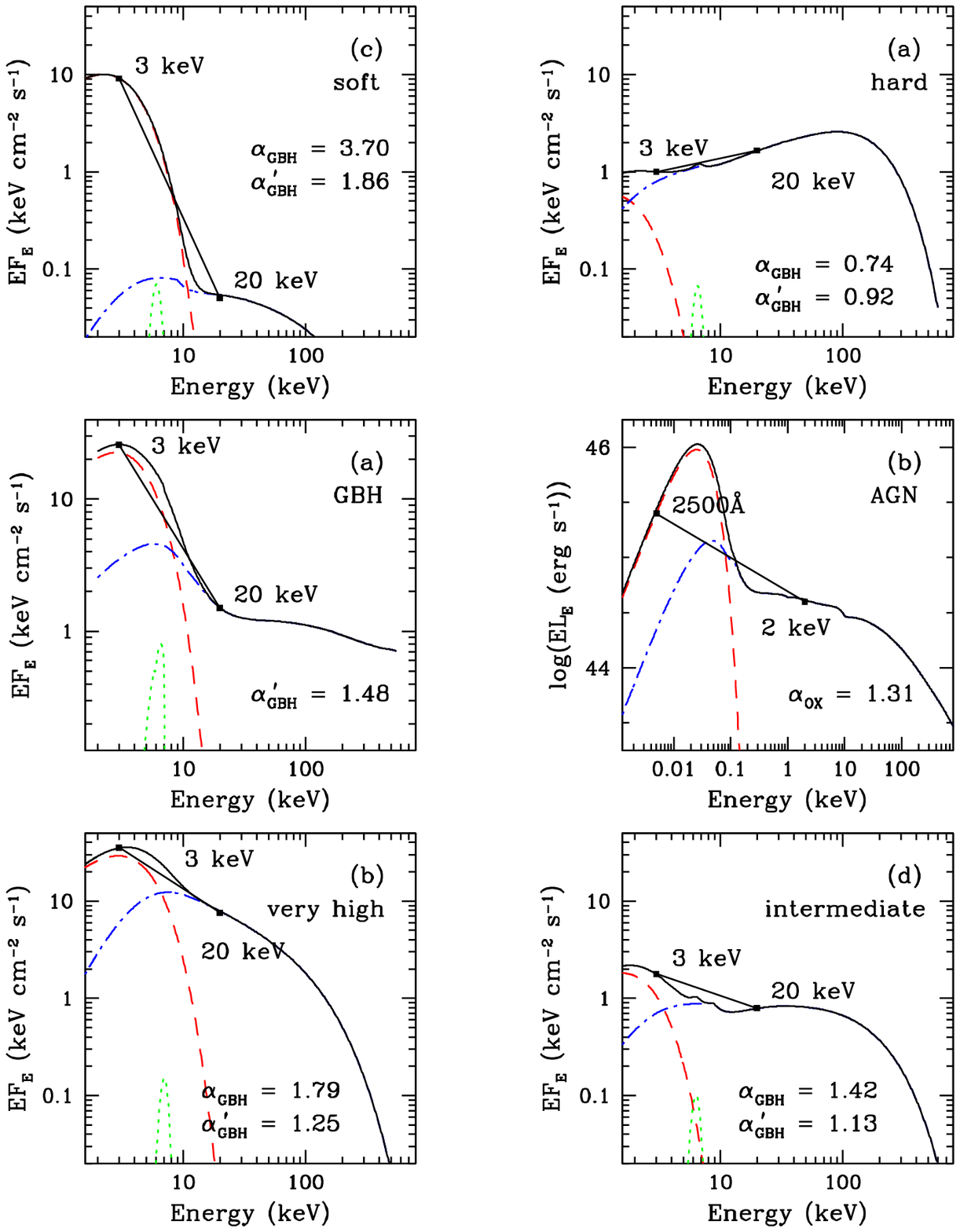}
\includegraphics[height=4.8cm,bb=50 150 196 325,clip]{fig2.ps}
\includegraphics[height=4.8cm,bb=50 500 196 675,clip]{fig2.ps}
\includegraphics[height=4.8cm,bb=280 150 425 325,clip]{fig2.ps}
\caption{Definition of the disc-to-Comptonization parameter in GBHs,
  $\alpha_{\rm GBH}$. (a) Disc blackbody (dashed/red) plus {\tt thcomp}
  (dot-dashed/blue) fit to a hard state GRO~J1655-40 spectrum. (b--d) Same
  model fit to a very high, soft and intermediate state GRO~J1655-40
  spectrum, respectively. There is also a small contribution from an iron
  $K_{\alpha}$ line (dotted/green). The spectra are corrected for the
  Galactic absorption. Spectra in (a--d) correspond to observations 2, 5, 6
  and 8 in Fig.~\protect{\ref{fig:fig3}}, respectively, described with 
  Model 1 (see text for detail of modeling and Sec.~\protect{\ref{sec:results}}
  for definition of the 'stretched' index $\alpha^{ \prime}_{\rm GBH}$.}.
\label{fig:fig2}
\end{figure*}

\begin{figure*}
\centering
\includegraphics[height=4.8cm,bb=50 327 320 500,clip]{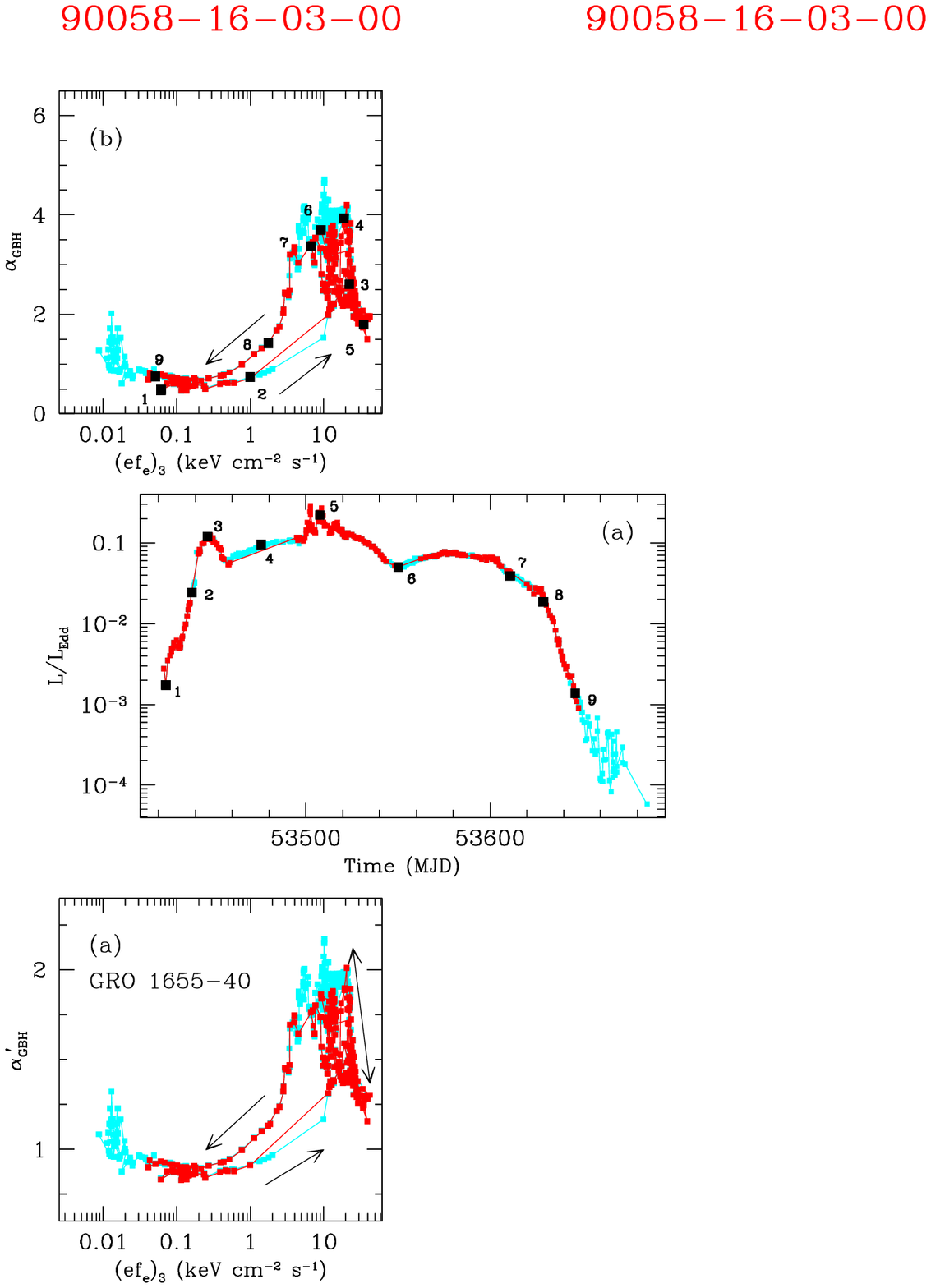}
\includegraphics[height=4.8cm,bb=20 500 195 673,clip]{fig3.ps}
\caption{(a) Light curve of GRO~1655--40 covering 260 days of the 2005
outburst. Evolution of the X-ray bolometric luminosity in Eddington
units is shown for 6.3 M$_{\odot}$ black hole 3.2 kpc away. See text
for details of 1-9. Method 1 approach is indicated by the light grey (cyan)
squares and Method 2 by dark grey (red) squares. (b) Dependence of
$\alpha_{\rm GBH}$ on $EF_{\rm E}$ flux at 3 keV for an outburst of
GRO~1655--40. Observations indicated with black squares correspond to
those in the left panel. Arrows illustrate how the source moves on the
diagram during an outburst.}
\label{fig:fig3}
\end{figure*}

In Fig.~\ref{fig:fig2} we show the shape of GRO~J1655-40 spectra in different
spectral states. The individual spectra correspond to observations 2, 5, 6
and 8 in the long term lightcurve of outburst presented in
Fig.~\ref{fig:fig3}a. Note that the luminosity of the source in these four
states is high in terms of the Eddington luminosity and it is always higher
than $L > 0.01 L_{\rm Edd}$. We calculated the Eddington luminosity for
isotropic emission from accreting black hole with mass of 6.3 M$_{\odot}$
located at the distance of 3.2 kpc \citep[][]{mccremi:03}. This outburst
contains all known spectral states: soft (including ultra-soft with very
weak hard X-ray tail, very high with strong hard X-ray emission, and
typical soft state dominated by disc emission and showing a power law with
soft photon index), intermediate and hard.

With the exception of the hard-state data, the spectra are dominated
by the disc at 3 keV. Conversely, hard X-ray power-law like radiation
dominates at 20 keV (see Sec.~\ref{sec:data} for details of the applied
model). Hence, for each GBH spectrum we find the counterpart of quasars'
parameter $\alpha_{\rm OX}$ 
(Eq.~\ref{eq:eq1}) defined as
\begin{equation}
\alpha_{\rm GBH} = -\frac{\log (ef_e)_{\rm 3} - \log (ef_e)_{\rm 20}}{\log
  3 - \log 20}+1,
\label{eq:eq2}
\end{equation}
where $(ef_e)_{\rm 3}$ and $(ef_e)_{\rm 20}$ are (unabsorbed) $EF_{\rm
E}$ (keV cm$^{-2}$ s$^{-1}$) fluxes at 3 and 20 keV, respectively.  This
definition is illustrated in Fig.~\ref{fig:fig2}. In
the hard state when the disc spectrum is cool and weak the
$\alpha_{\rm GBH}$ measures effectively the photon index of the
Comptonized component and may not correspond directly to the
$\alpha_{\rm OX}$ in AGN (Fig.~\ref{fig:fig2}a).


\section{Data reduction and analysis}
\label{sec:data_ra}


\subsection{Data}
\label{sec:data}

We analysed {\it RXTE} data from a number of Galactic black hole
X-ray binaries: GRO~J1655--40, XTE~J1550--564, XTE~J1650--500,
XTE~J1859+226, H1743+332 and GX~339--4. We used all archive data available
before June 2007, in total 1930 pointed observations. The best
outburst coverage is available for GRO~J1655--40, XTE~J1550--564
(1998/1999 outburst) and XTE~J1650--500. Table \ref{tab:tab1} contains list
of sources, time-span of outbursts and observed spectral states.

We reduced data from public HEASARC\footnote{High Energy
Astrophysics Science Archive Research Center} archives using {\sc
ftools} ver. 6.2. We extracted PCA\footnote{Proportional Counter
Array} Standard 2 spectra for detector 2, top layer only, in the 3--20 keV
band  and HEXTE spectra from both detectors in the 20--200 keV band. We
obtained one spectrum per pointed observation and we accounted for systematic errors in PCA data at the level of 1\%.
For background estimate we used the latest background models available from {\it RXTE} data analysis web-pages, together with appropriate background model files. To enhance the signal-to-noise ratio, $S/N$, we rebinned the PCA and HEXTE spectra requiring that $S/N>5$ in each new bin (which corresponds to $\ga$20 source photons in each new bin). We modeled the spectra
using {\sc xspec} ver. 11.3 \citep[][]{arnaud:1996} to find the photon index, and the $EF_{\rm E}$ fluxes at 3 and
20 keV. Based on these monochromatic fluxes we estimated the relative strength
of the disc and hard X-ray fluxes parametrized by $\alpha_{\rm GBH}$ (Eq.~\ref{eq:eq2}).
We applied two methods of data analysis.

In the first method (Method 1), we fitted simultaneously the PCA/HEXTE
(3--200 keV) spectra with a model consisting of a multicolour disc
blackbody, thermal Comptonization \citep*[{\sc thcomp} in {\sc
xspec};][]{zdziarski:96}, Gaussian line profile to model iron
K$\alpha$ feature, and smeared edge ({\sc smedge}) to account for iron
K$\alpha$ absorption. We also added Galactic absorption with $N_{\rm H}$
fixed throughout the outburst at best values from literature (Table~\ref{tab:tab1}), and a constant
to account for differences in normalization between PCA and HEXTE (the
constant has been fixed at 1 for PCA). The complete model is described
in {\sc xspec} as {\sc constant*wabs*smedge*(gaussian + diskbb +
thcomp)} \citep[see][ for details]{donegier:03}. Several data sets
show evidence of a  non-thermal tail at high energies \citep[compare, e.g.,][]{zdziarskiea:2001}, which was not accounted
for by the model. However, the discrepancy is significant only at the energies 
higher than 20 keV. In PCA bandpass (in particular at 3 and 20 keV) the
model describes the data well, and the reduced $\chi_{\nu}^2$ for the fit
to the full PCA/HEXTE band is always lower than 2. We determine the $(ef_e)_{\rm 3}$ and $(ef_e)_{\rm 20}$ fluxes directly
from these spectral fits.


\begin{figure*}
\centering
\includegraphics[height=3.4cm,bb=30 175 195 320,clip]{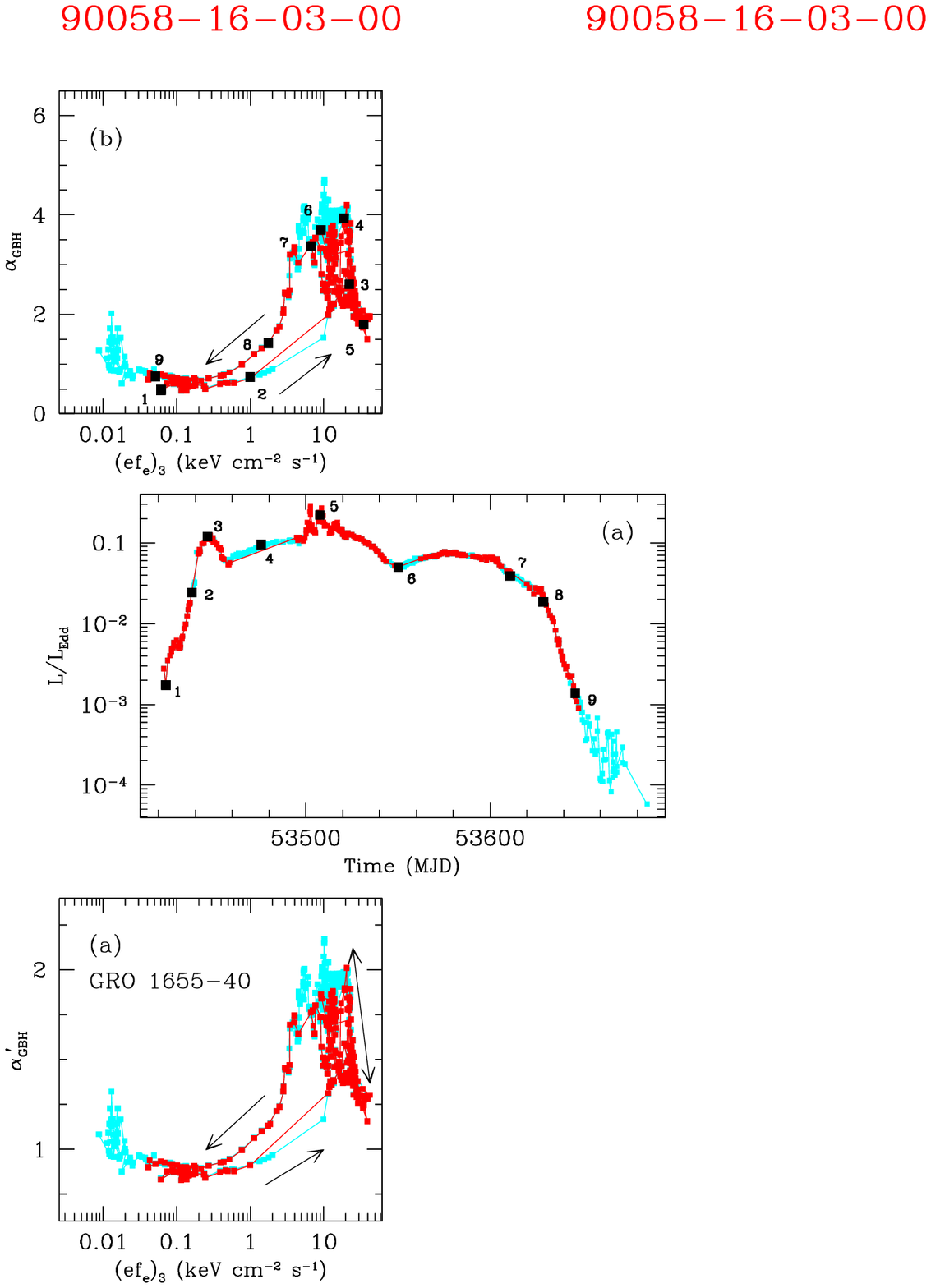}
\includegraphics[height=3.4cm,bb=52 175 195 320,clip]{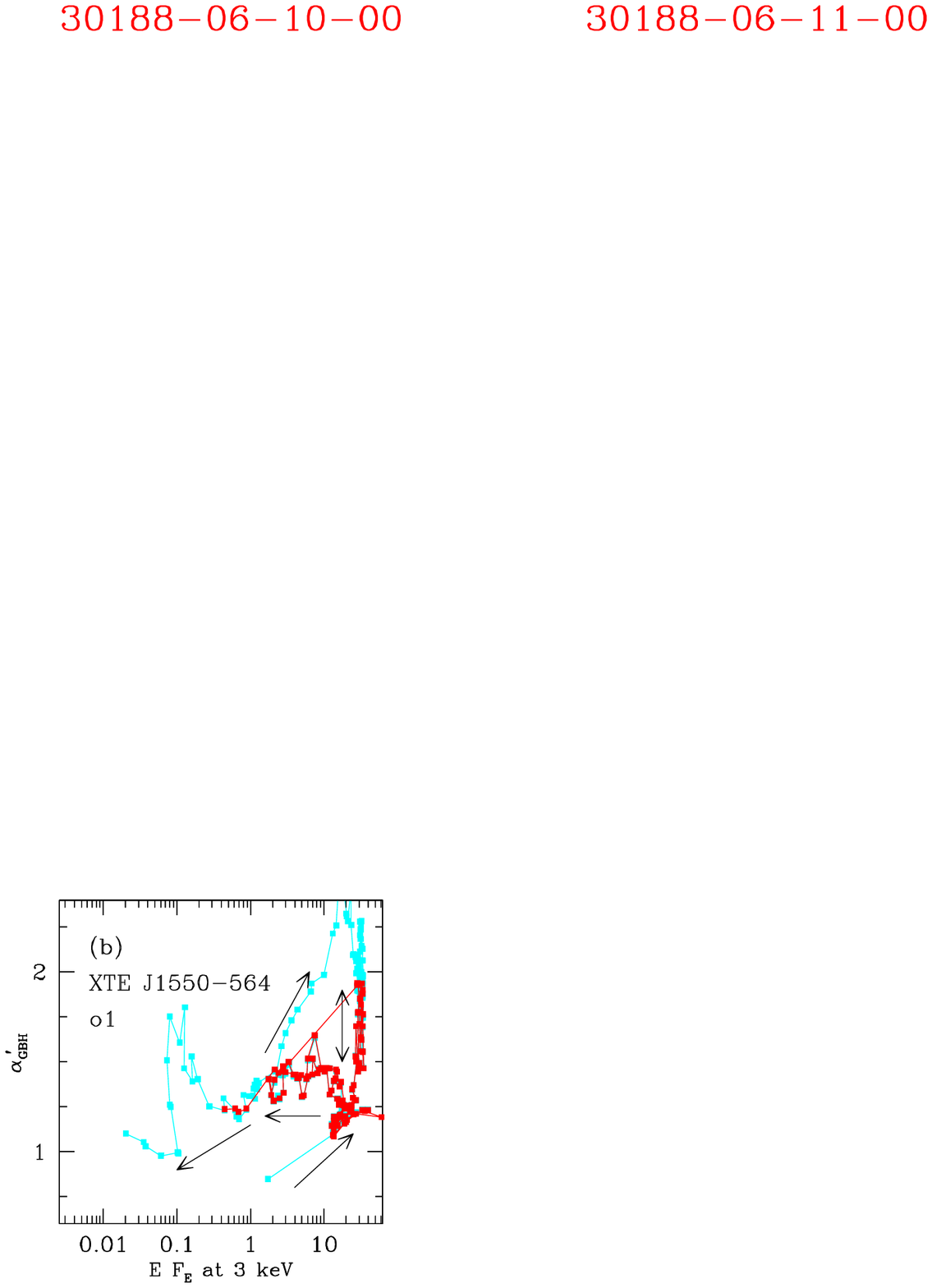}
\includegraphics[height=3.4cm,bb=52 175 195 320,clip]{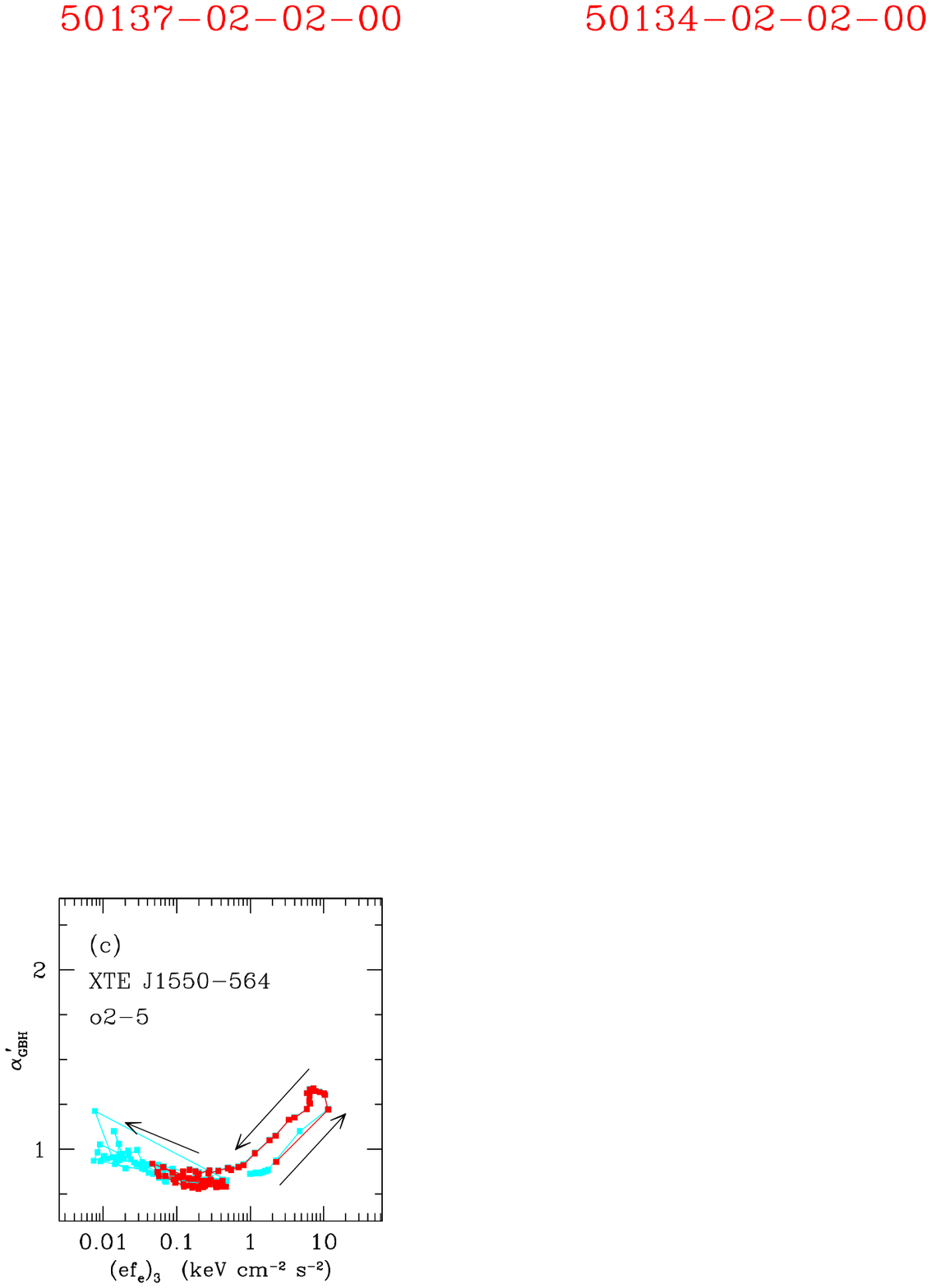}
\includegraphics[height=3.4cm,bb=52 175 195 320,clip]{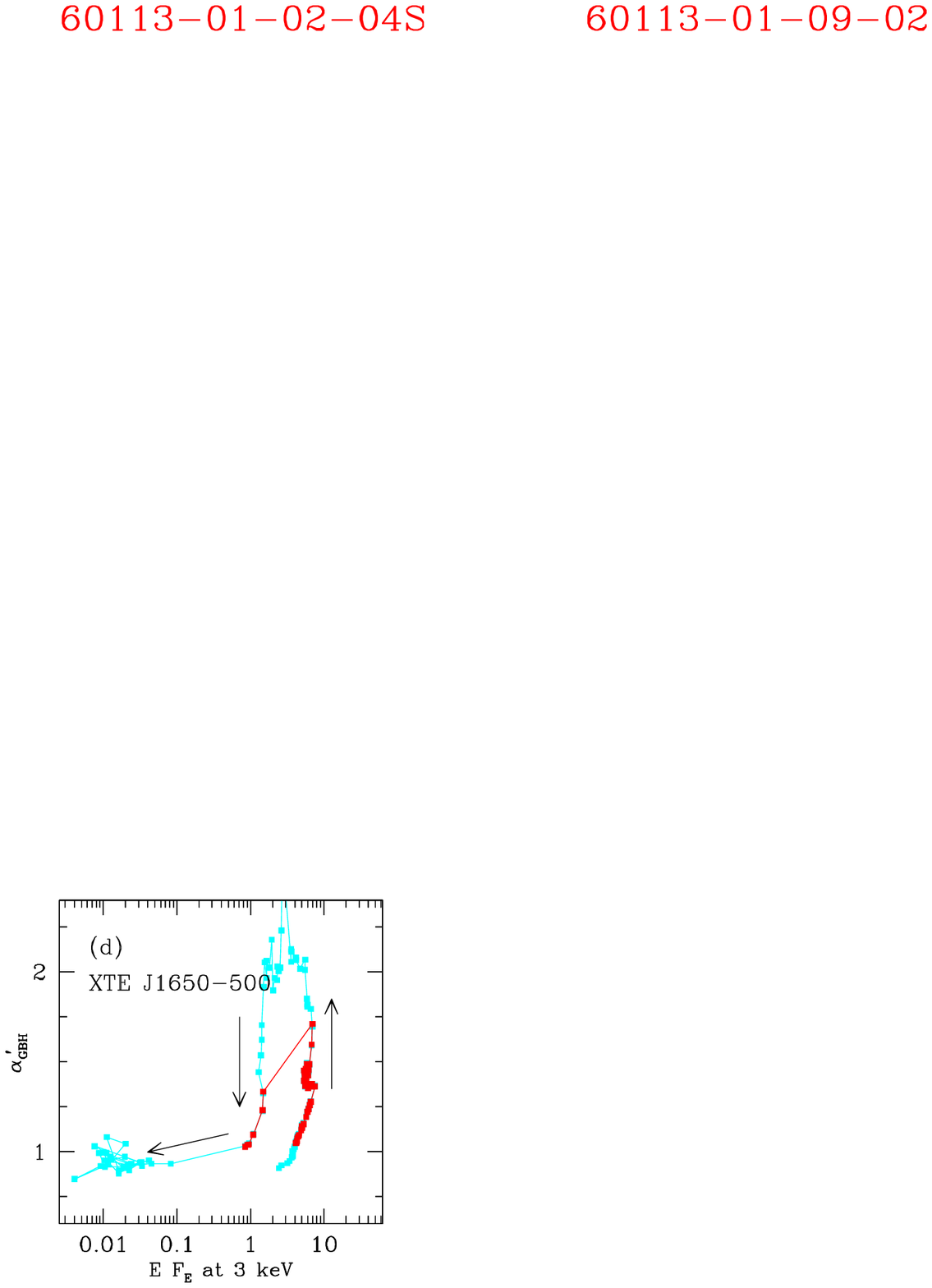}
\includegraphics[height=3.4cm,bb=52 175 195 320,clip]{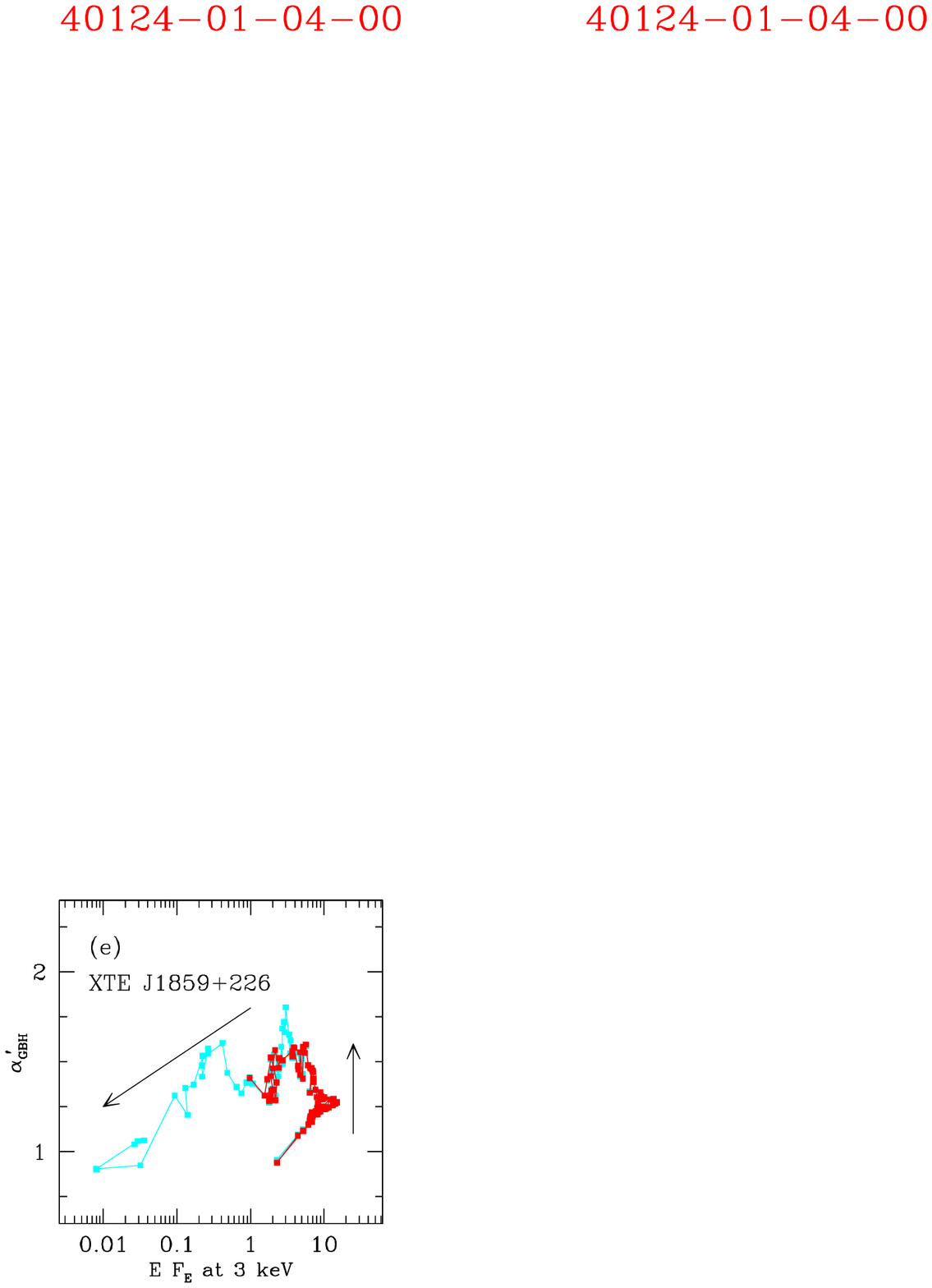}
\includegraphics[height=4cm,bb=30 150 195 320,clip]{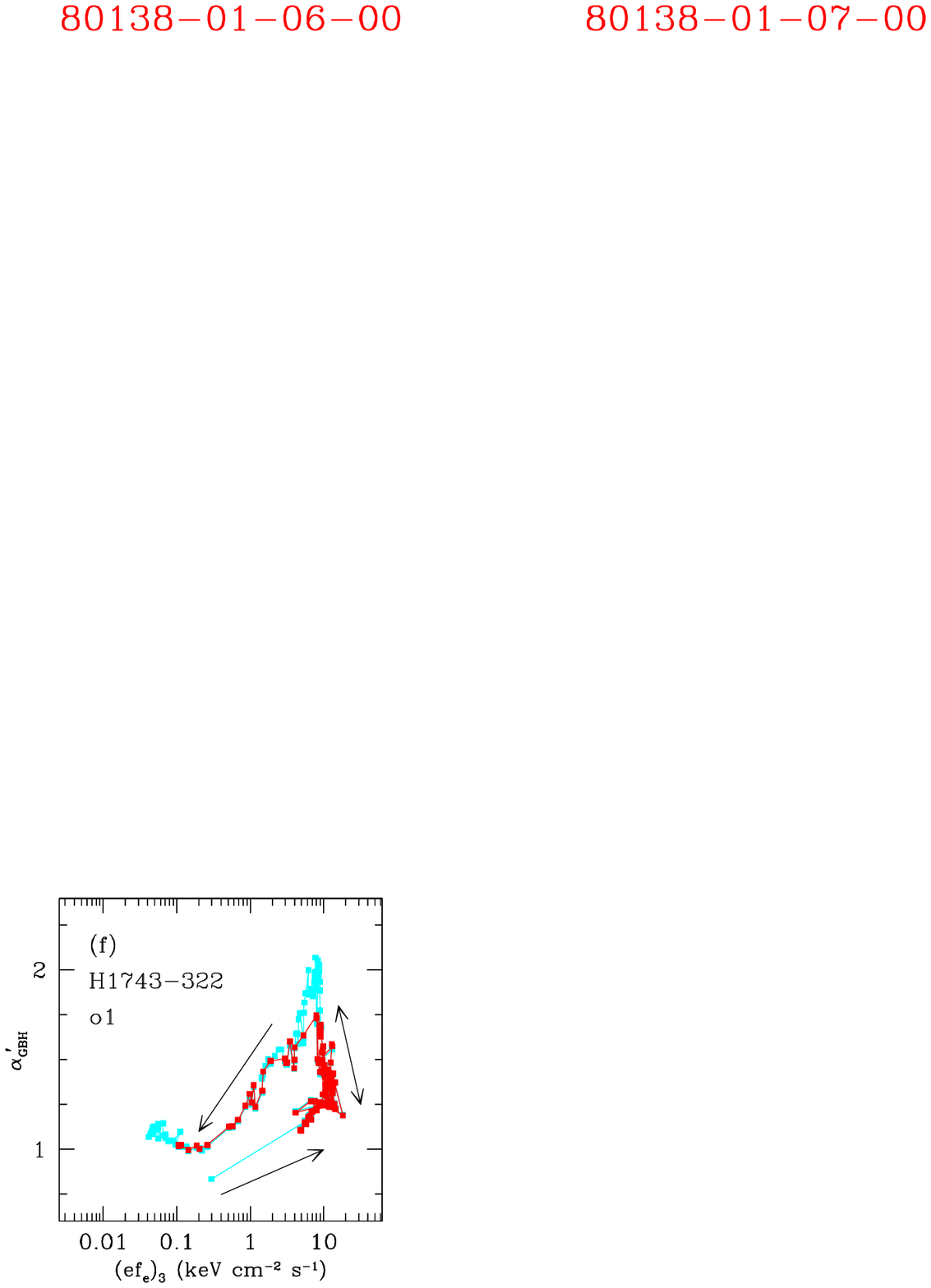}
\includegraphics[height=4cm,bb=52 150 195 320,clip]{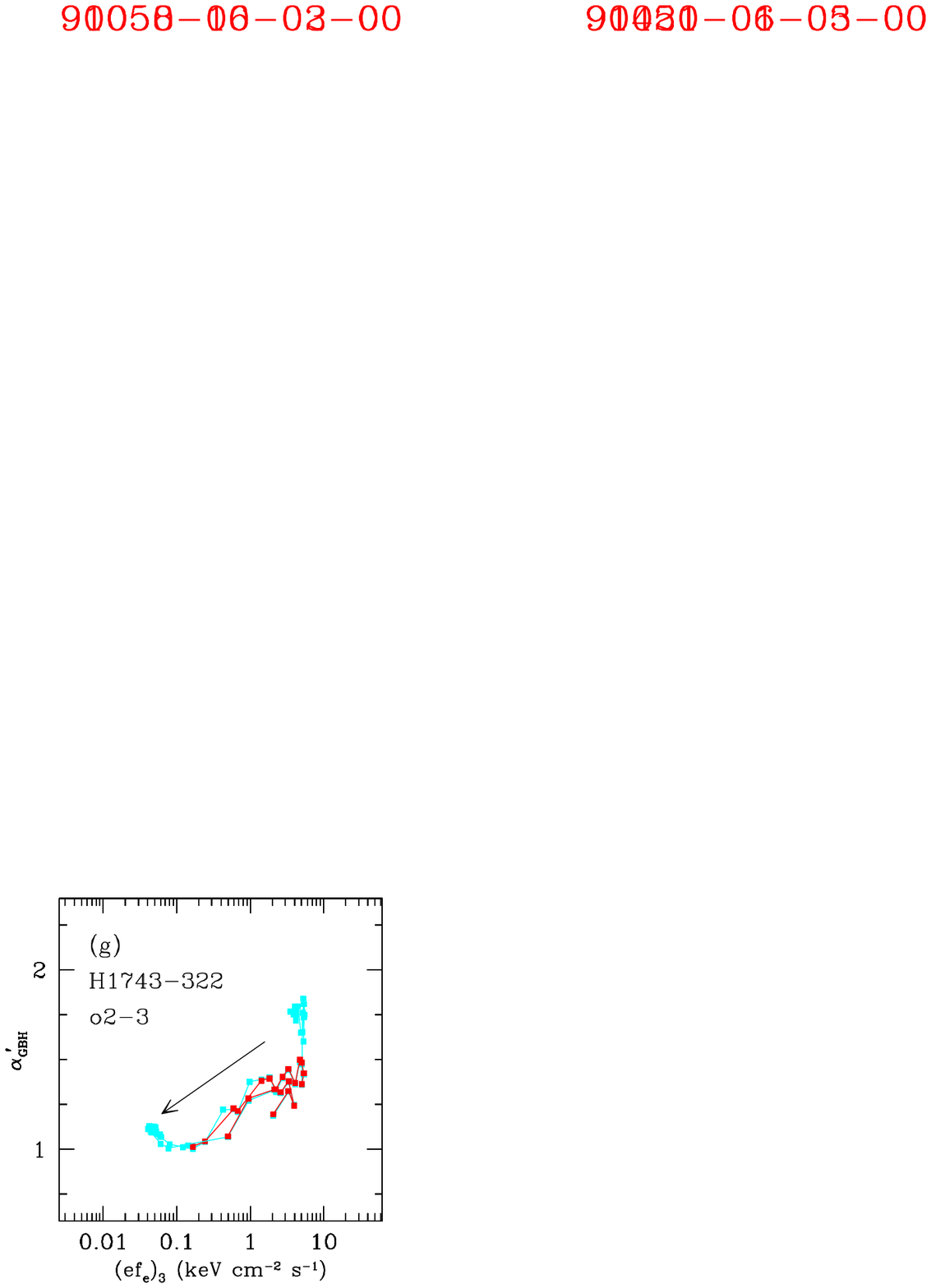}
\includegraphics[height=4cm,bb=52 150 195 320,clip]{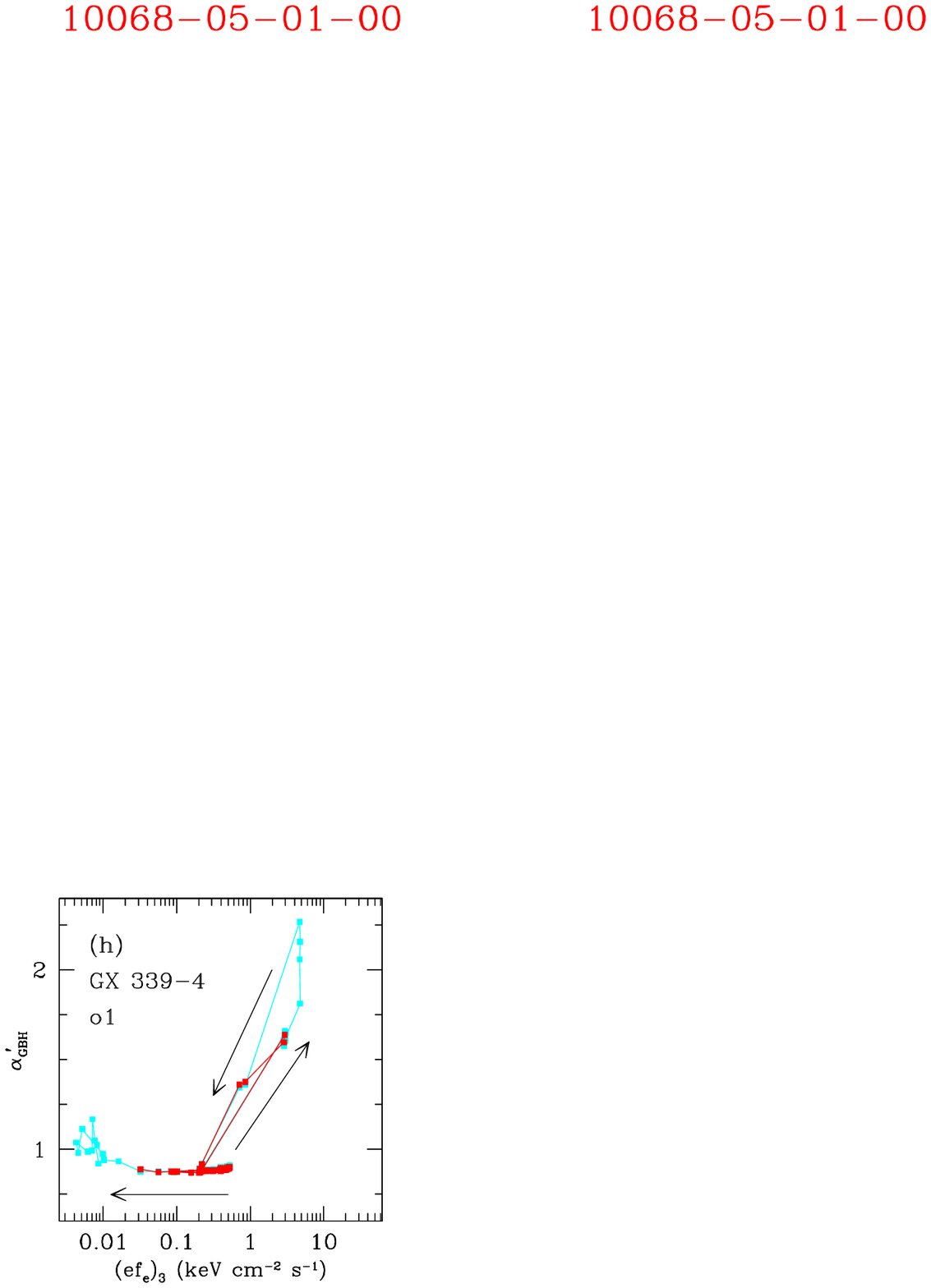}
\includegraphics[height=4cm,bb=52 150 195 320,clip]{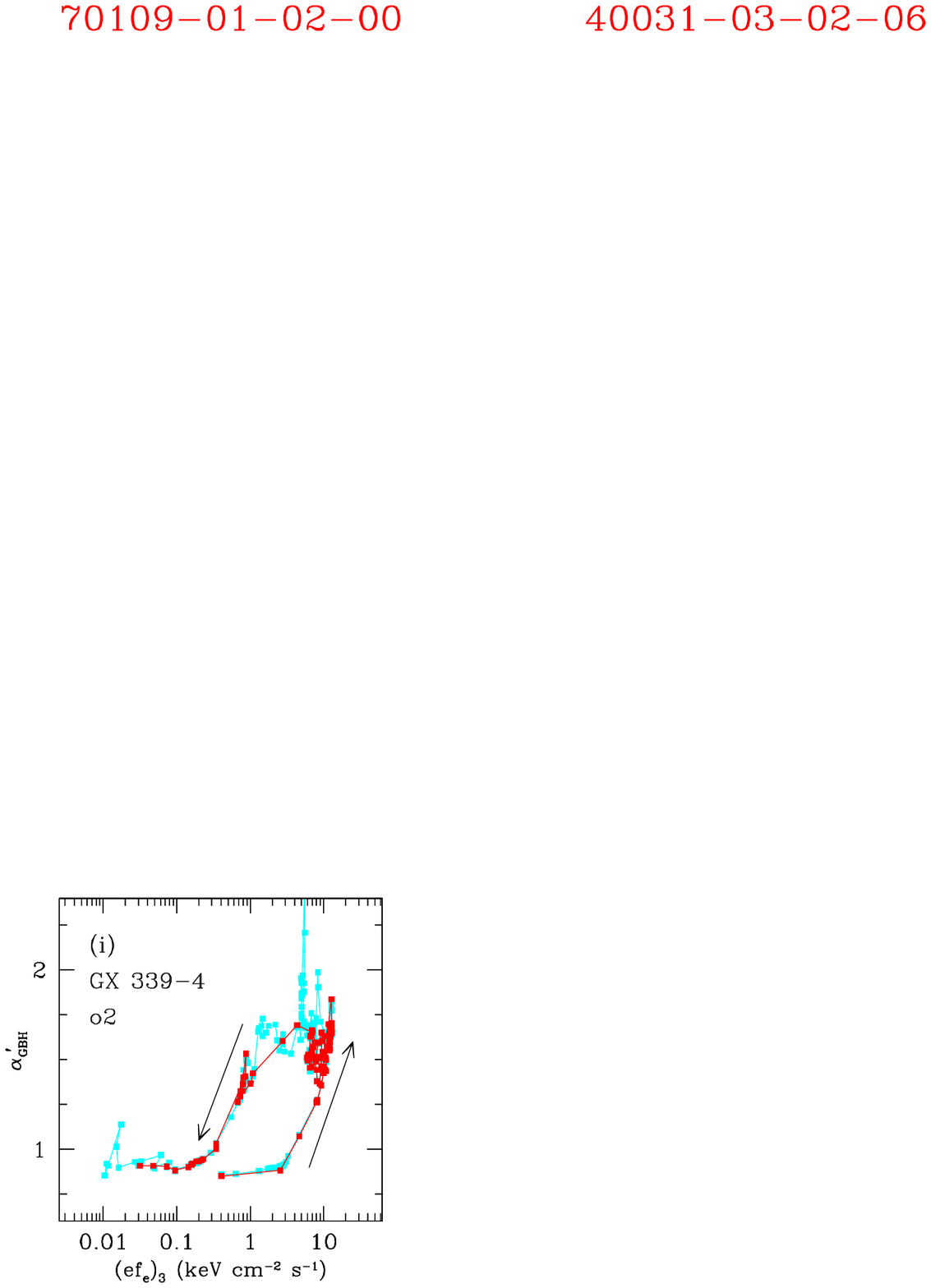}
\includegraphics[height=4cm,bb=52 150 195 320,clip]{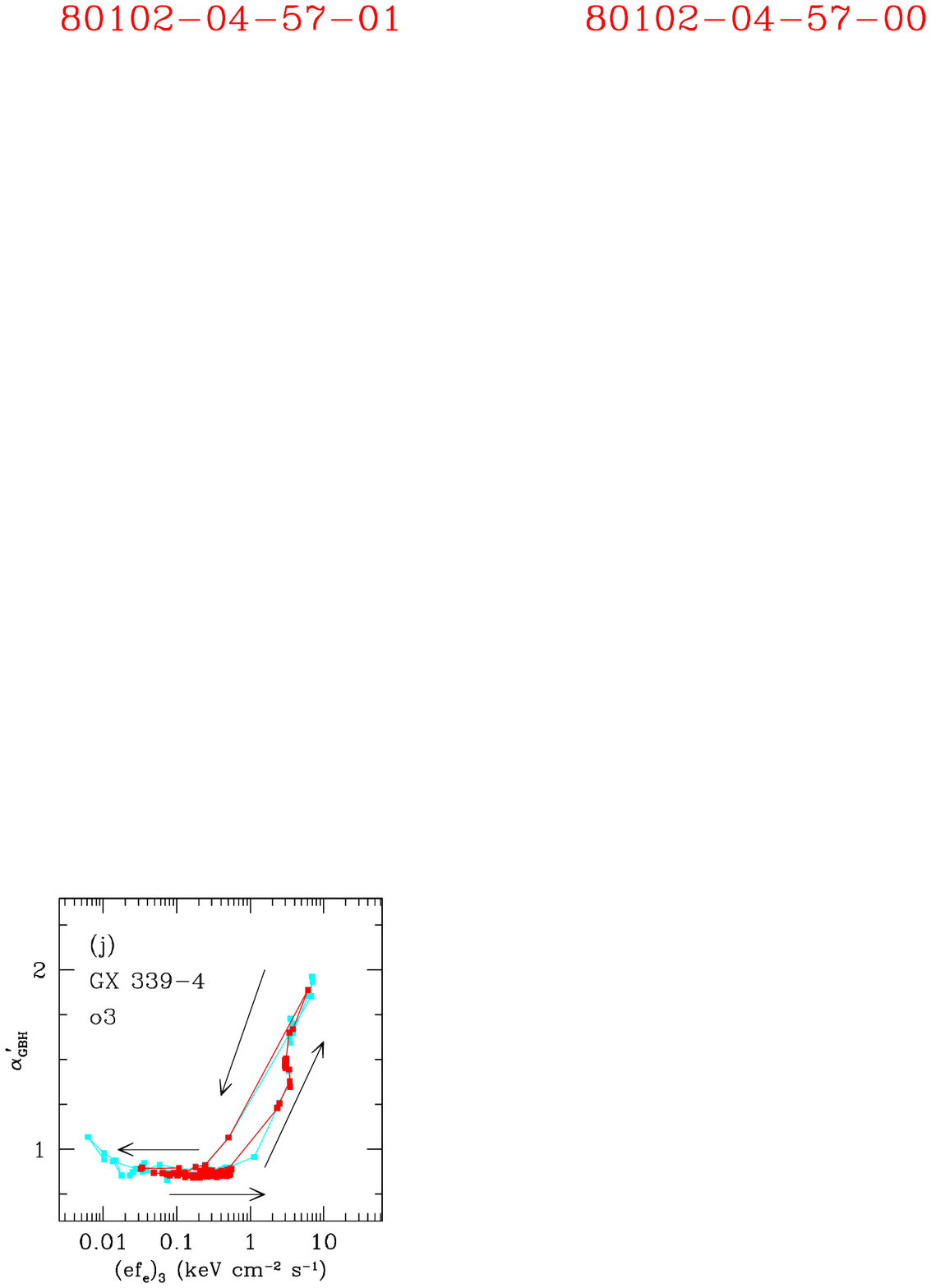}
\caption{Dependence of $\alpha^{ \prime}_{\rm GBH}$ on the monochromatic 3
keV flux for a number of Galactic black hole X-ray binaries.
Typically, $1 < \alpha^{ \prime}_{\rm GBH} < 2$, as in AGN. The results of
Method 1 (based on the fits to the broad band 3--200 keV
PCA/HEXTE data) and Method 2 (based on the fits to the 20--40 keV
HEXTE data) are compared and indicated with light grey (cyan) and dark grey (red),
respectively. The arrows indicate direction of movement in the
diagrams during an outburst. For XTE~J1550--564, H1743--332 and
GX~339--4 we plot consecutive outbursts separately for clarity of
presentation.}
\label{fig:fig4}
\end{figure*}


\begin{figure*}
\centering
\includegraphics[width=5.7cm,bb=18 144 565 675,clip]{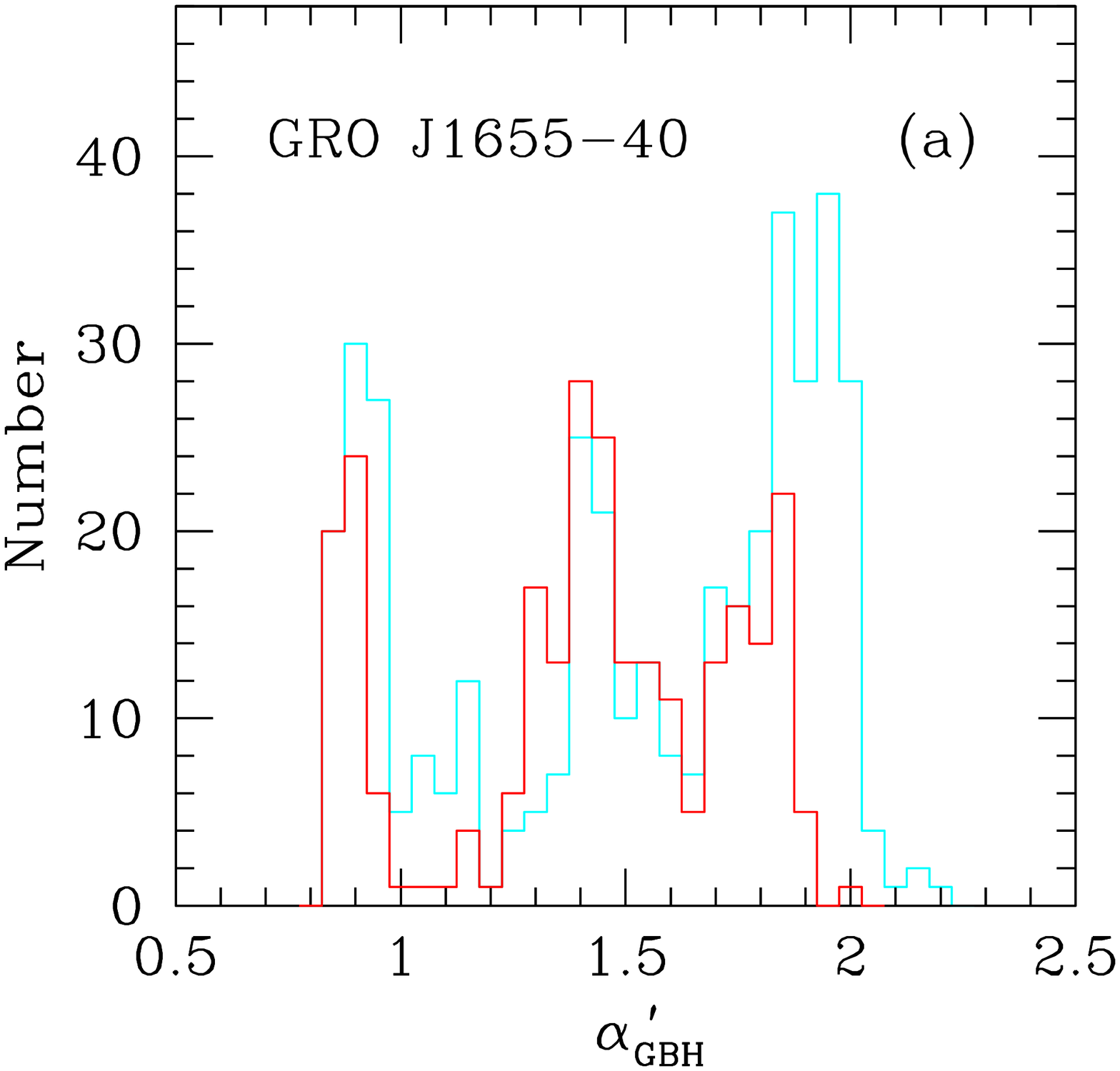}
\includegraphics[width=5.7cm,bb=18 144 565 675,clip]{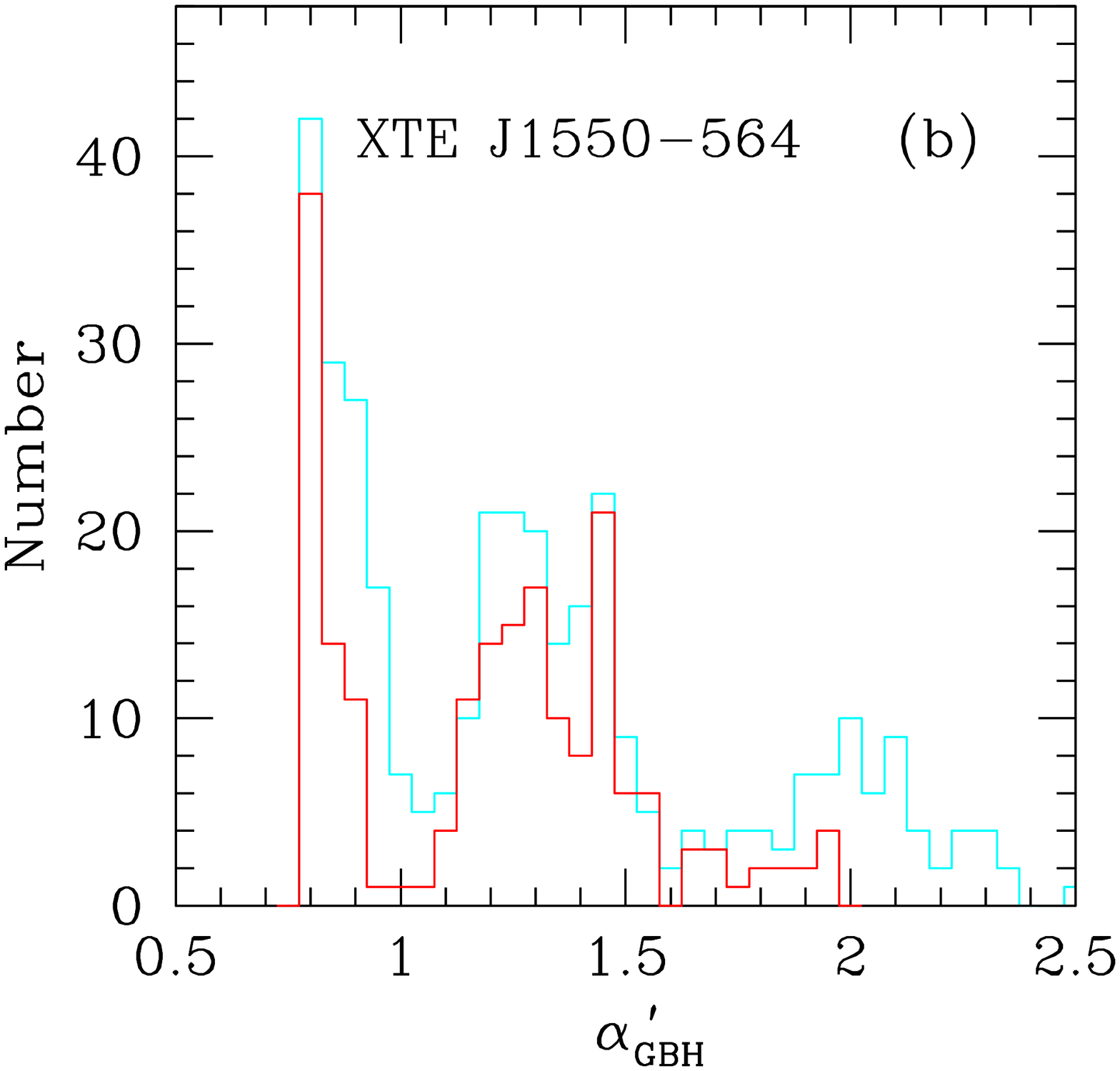}
\includegraphics[width=5.7cm,bb=18 144 565 675,clip]{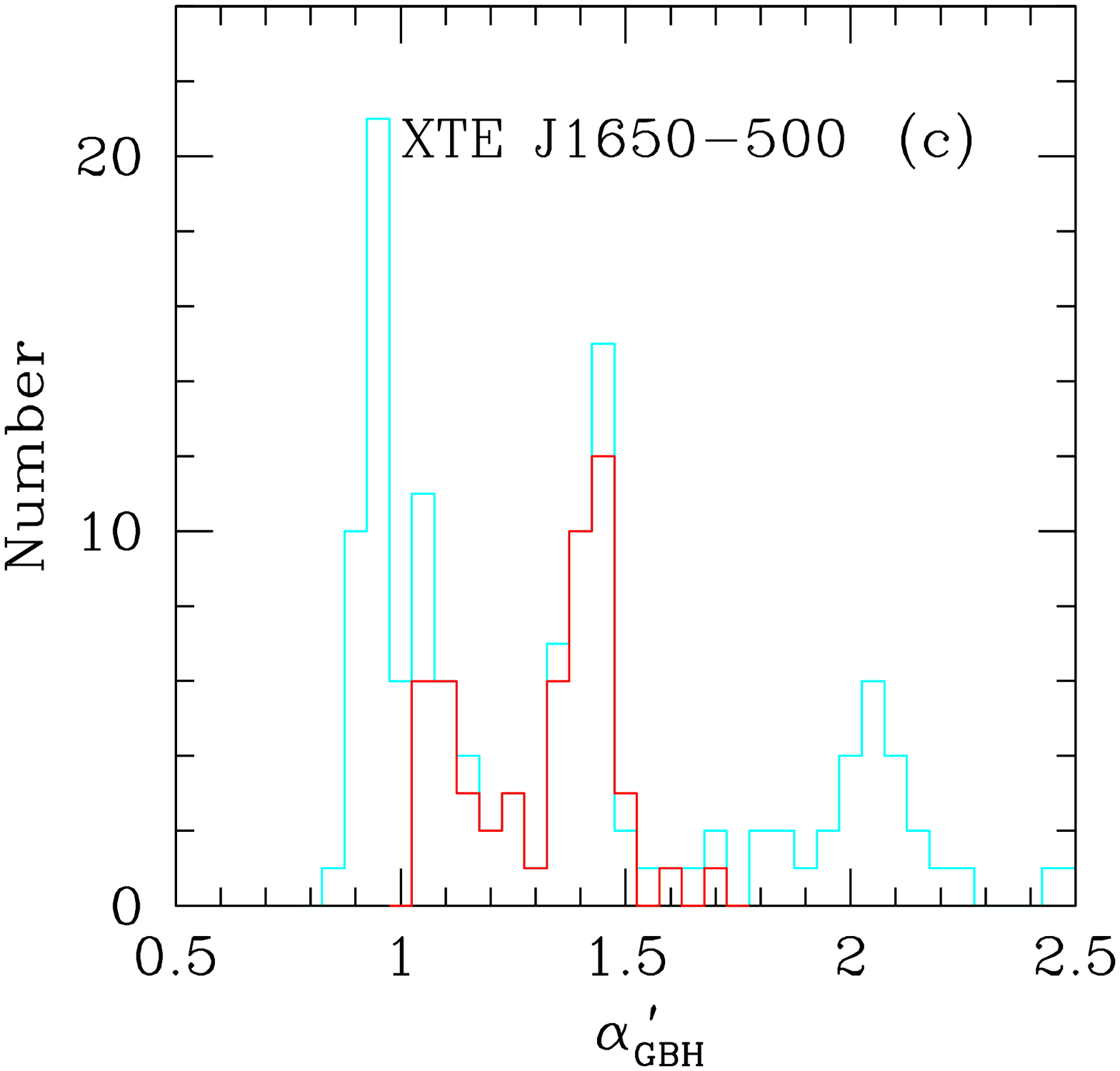}
\caption{Distribution of $\alpha_{\rm GBH}^{~\prime}$ for GBHs. Three main
  peaks can be distinguished close to $\alpha_{\rm GBH}^{~\prime}$ values
  of 
1, 1.5 and 2. These peaks correspond to a hard, very high/intermediate
and typical soft/ultra-soft spectral states, respectively. The light grey
(cyan) 
histograms are based on Method 1 (fits to the 2--200 keV PCA/HEXTE
data), while the dark grey (red) histograms are based on Method 2 (fits to the
20--40 keV HEXTE data).  It can be seen that in Method 2 the peak
corresponding to the hard state (centred around 1) gets significantly
diminished. Similarly, the soft state peak (centred at ${\sim}2$) may
be diminished or may even disappear from the distribution.}
\label{fig:fig5}
\end{figure*}


We use only these data sets that give reduced $\chi_{\nu}^2 < 2$, in total 1691
observations. Among them, some soft state observations that are bright in the PCA bandpass, show very low HEXTE flux.
In addition, the hard state data at the beginning and end of an outburst show overall low flux in both the PCA and HEXTE bandpasses.
In these two cases we estimate $\alpha_{\rm GBH}$ based on PCA data, and use only these PCA data sets that after rebinning contain at least 10 bins with $S/N>5$.

In the second method (Method 2), we consider the 20--40~keV HEXTE
hard X-ray band with negligible contribution from the disc. We
perform {\sc constant*powerlaw} fits with the constants for each HEXTE detector fixed at the values found in Method 1. We used only the data sets which gave the reduced $\chi^2 < 2$ and for which the photon index could be constrained between 1 and 8 at 90 per cent confidence level, in total 1006 observations. In this method, the $(ef_e)_{\rm 20}$ is derived from 20--40 keV HEXTE fits and $(ef_e)_{\rm 3}$ is taken from models described in Method 1.

In both methods the photon index is a fit parameter and we use Eq.~\ref{eq:eq1} to calculate $\alpha_{\rm GBH}$.
Method 1 is more comprehensive because in general it is based on
the broad band PCA/HEXTE fits. Hence, the results drawn from Method 1 can
be regarded as a reference for our considerations. Conversely, Method 2 `mimics' the
methodology commonly applied to the AGN, for which conclusions
about radiation mechanisms are drawn based on incomplete information about the spectra, i.e. discontinuous observations taken in optical/UV (cool disc)
and X-rays. Thus, results of Method 2 are better suited for GBHs vs. AGN comparison.

Studies of large samples consisting of AGN observed with different X-ray missions
cannot rely on hardness ratios because this would introduce instrumental effects 
Instead, $\alpha_{\rm OX}$ parametrisation based on spectral fits is used. 
We also use spectral fits to perform a consistent analogy between the GBHs and AGN. We note however that we get consistent
and robust results on $\alpha_{\rm GBH}$ independently on the
spectral model as long as the fit is good (see Sobolewska, Gierli\'{n}ski,
Siemiginowska 2009 in preparation, where we use {\tt eqpair} to fit the data of GRO~J1655-40).
In the remaining of the paper we compare GBHs with the AGN. We investigate any possible
issues in the interpretation of the AGN data, determine the spectral
state of the observed AGN, and identify classes of objects that may be
missing.

\subsection{Example of GRO~J1655-40}

Figure~\ref{fig:fig3}b shows $\alpha_{\rm GBH}$ as a function of
monochromatic disc flux at 3 keV, $(ef_e)_{\rm 3}$, for GRO~J1655-40. We
indicate the results of Methods 1 and 2 with cyan 
and red colours, respectively. There is a characteristic circular
pattern in the diagram resembling that seen in the GBHs hardness-intensity
diagrams \citep[e.g.][]{fenderea:2004,dunnea:2008}. However,
hardness-intensity diagrams in the literature use diverse definitions of
hardness, and based on them it is difficult to draw direct analogies
between GBHs and AGN. Uniform parametrization with $\alpha_{\rm GBH}$ that we
propose in this paper makes such a comparison straightforward (see
Sec.~\ref{sec:results}). Below we describe how the spectral evolution of
GRO~J1655-40 throughout an outburst affects evolution of $\alpha_{\rm
  GBH}$. This can be generalized to the case of other GBHs, as we
show in Sec.~\ref{sec:results}. 

There is no linear correlation between the $\alpha_{\rm
GBH}$ and $(ef_e)_3$, and there is no unique value of $\alpha_{\rm GBH}$
characteristic of a given spectral state. At the beginning of the outburst
(observation 1 in Fig.~\ref{fig:fig3}) 
the source is in a hard state, the $\alpha_{\rm GBH}$ is less than unity
and it simply provides a measure of the slope of the Comptonization. With time
disc and Comptonization luminosities increase, but the spectral shape does not change much, and
$\alpha_{\rm GBH}$ stays approximately constant (observation 2). During the
hard-to-soft state transition the source moves toward the 
upper right corner of the plot (observation 3) and $\alpha_{\rm GBH}$
increases. In the soft spectral state the disc flux does not vary
significantly as opposed to the Comptonized component flux, and wide range
of $\alpha_{\rm GBH}$ is covered while the source moves up and down on the
diagram (observations 3, 4, 5 and 6). Observation 5 is an example of the
very high state spectrum, resulting in moderate $\alpha_{\rm GBH}$
and position in the lower right corner of the diagram. In the ultra-soft
state $\alpha_{\rm GBH}$ reaches the maximum (observation 6). Observation 7
marks the beginning of the soft-to-hard state transition. The $\alpha_{\rm
  GBH}$ decreases  (observation 8) and the source moves back to 
the hard state (observation 9). 

Data succeeding observation 9 have statistics too poor to be considered in
Method 2, because HEXTE is not sensitive enough to give reasonable
results for such low luminosity hard states. Similarly, a number of data
sets surrounding observation 4 were excluded by Method 2. These data
represent an ultra-soft state, with a weak Comptonized tail. Future
observations with more sensitive hard X-ray detectors  (e.g. NuStar, EXIST)
will provide a better coverage for the sources in the low luminosity hard
state. 


\section{Results: Comparison of GBH binaries and AGN}
\label{sec:results}

\subsection{Distribution of the disc-to-Comptonization index}

Central black holes in AGN are 5--8 orders of magnitude more massive
than in GBHs. Disc temperature scales with mass as $M^{-1/4}$
for a given fraction of Eddington luminosity, so the disc is cooler in
AGN and emits in optical/UV band instead of soft X-rays. For AGN, the
disc-to-Comptonization slope called X-ray loudness is defined
for $2500\AA\sim 0.005$ keV and 2 keV, whereas we define our
$\alpha_{\rm GBH}$ for 3 and 20 keV.  This means that the
disc-to-Comptonization slope, parametrized by $\alpha_{\rm GBH}$ in
GBHs, needs to be scaled, or `stretched', before it can be directly
compared with $\alpha_{\rm OX}$ in AGN. We account for the shift of
the disc spectrum towards lower energies in AGN by defining
\begin{equation}
\label{eq:eq3}
\alpha^{ \prime}_{\rm GBH} = C(\alpha_{\rm GBH}-1)+1,
\end{equation}
where $C = (\log 3-\log 20)/(\log 0.005 -\log 2) \approx 0.32$, and
$\alpha_{\rm GBH}$ is given by Eq.~\ref{eq:eq2}. After this
correction, a spectrum of a GBH and a spectrum of an AGN with the same
ratio of the disc normalization to the Comptonization normalization
will result in a comparable $\alpha^{ \prime}_{\rm GBH}$ and
$\alpha_{\rm OX}$ indices.

We have found $\alpha^{ \prime}_{\rm GBH}$ for a number of GBHs.
In Fig.~\ref{fig:fig4} we show the results of calculating
$\alpha^{ \prime}_{\rm GBH}$ both directly from the PCA/HEXTE fits (Method
1, cyan) and from the method that mimics the AGN studies, and
uses only a part of the hard X-ray spectrum, here 20--40 keV
(Method 2, red). Both methods give similar results, but Method 2
is more restrictive and excludes significant number of data points in the
soft and ultra-soft states and in the hard state with low luminosity, where
the HEXTE data are too weak to constrain $(ef_e)_{\rm 20}$.
Figure~\ref{fig:fig4} shows that the shape of the track in the $\alpha_{\rm GBH}$ (and hence $\alpha^{
  \prime}_{\rm GBH}$) and $(ef_e)_{\rm 3}$ diagram found 
for GRO~J1655--40 (Fig. \ref{fig:fig3}) is similar in other
sources.

Scaling disc-to-Comptonization index according to
Eq.~\ref{eq:eq3} results in $\alpha^{ \prime}_{\rm GBH}$ parameter ranging
between $\sim$1 and $\sim$2, which is consistent with the AGN
observations as shown by histogram of $\alpha_{\rm OX}$
(Fig.~\ref{fig:fig1}) produced for the optically selected
radio-quiet quasars cross-correlated with the Chandra X-ray archive
\citep[][]{kelly:2007} and for radio-loud {\it ROSAT} quasars \citep[][]{greenea:1995}. For the AGN samples, $\alpha_{\rm OX}$
distributions peak at 1.5 and 1.3, respectively. To compare the behaviour of AGN
and GBHs we produced histograms of $\alpha^{ \prime}_{\rm GBH}$
for three Galactic sources with the best coverage of
their outbursts: GRO~1655--40, XTE~J1550--564 and XTE~J1650--500
(Figs.~\ref{fig:fig5}a--c). In the case of XTE~J1550--564, we considered
all its  5 outbursts together. Three main peaks can be
distinguished in the GBHs distributions, close to $\alpha^{ \prime}_{\rm GBH}$
of 1, 1.5 and 2. These peaks correspond to the hard, very
high/intermediate and typical-soft/ultra-soft spectral states,
respectively. (Shapes of the very high and intermediate states
energy spectra are very similar, with the only difference being
much higher luminosity in the very high state.) A comparison of
Figs.~\ref{fig:fig5}a--c and Fig.~\ref{fig:fig1} for GBHs and
AGN, respectively, suggests that majority of the observed AGN are
most probably in a spectral state corresponding to the very
high/intermediate state of GBHs.

We stress the difference between the $\alpha^{ \prime}_{\rm GBH}$
distributions arising from Methods 1 (cyan) and 2 (red). When Method 2
is applied (the one that corresponds to AGN methodology), the soft and hard
state peaks (centred at $\alpha^{\prime}_{\rm GBH} \simeq 2$~and~1,
respectively) are suppressed, which is most clearly visible in the case of XTE~J1650-500 for which the soft
state peak disappears from the distribution (Fig.~\ref{fig:fig5}c).

In the
remaining part of this section we use only the $\alpha^{ \prime}_{\rm GBH}$
and $\Gamma$ resulting from Method 2 to better compare the GBHs and AGN.


\subsection{Disc-to-Comptonization index vs. the photon index}


\begin{figure}
\centering
\includegraphics[height=8cm]{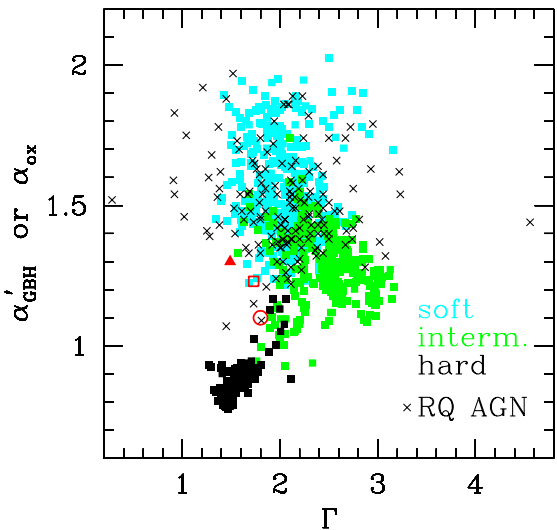}
\caption{Relation between the photon index, $\Gamma$, and
  disc-to-Comptonization index: $\alpha_{\rm OX}$ for radio-quiet quasars
  \protect{\citet[black crosses,][]{kelly:2007}} and $\alpha^{ \prime}_{\rm GBH}$ for GBHs in a hard
  (black), intermediate (dark grey/green) and soft (light grey/cyan)
  spectral state. We plot data for all GBHs considered in this
  paper. Average properties for radio-loud AGN are derived for samples of
  \protect{\citet[][ filled triangle]{elvisea:1994}} and \protect{\citet[][ open
  square]{gambillea:2003}}. We also show an average for a sample of LINERs from \protect{\citet[][ open circle]{maoz:2007}}.}
\label{fig:fig6}
\end{figure}


In Fig.~\ref{fig:fig6} we plot the 'stretched' disc-to-Comptonization
index, $\alpha^{\prime}_{\rm GBH}$, versus the photon spectral index,
$\Gamma$. The light grey/cyan, dark grey/green and black squares show
different spectral 
states of GBHs: soft, intermediate and hard, respectively. Note that
our soft state includes in addition to a typical soft state
also the ultra-soft (with a very weak X-ray tail) and very high (with
a strong disc and strong Comptonization) state spectra. The data form an
S-like track in the diagram. In the hard state the photon index is
hard with a median $\Gamma = 1.53$, the disc flux is weak, so the
median $\alpha^{\prime}_{\rm GBH} = 0.87$ is low, and the data occupy the
lower left corner of the diagram. Upper part of the diagram
corresponds to the soft state GBHs. It is worth noting that $\alpha^{\prime}_{\rm GBH}=2$ is most
probably not the real upper limit for $\alpha^{\prime}_{\rm GBH}$, but it
results from HEXTE sensitivity limits to detect 20--40 keV flux. The centre of the diagram is occupied by the very high/intermediate state GBHs.

We plot the Type 1 radio-quiet quasars data \citep[black crosses,][]{kelly:2007} with black crosses in the same figure and it can be seen that majority of them cluster in the centre of the diagram, and so they coincide with the very high/intermediate state GBHs, while minority with $\alpha_{\rm OX}$ greater than 1.6--1.7 match the soft state GBHs. All but four of these AGN have
photon index smaller than $\Gamma <3$ and $\alpha_{\rm OX}$ 
between 1 and 2. This sample does not contain quasars with a very
soft photon index. However, some Narrow Line Seyfert 1 galaxies have
$\Gamma>3$ \citep{boller} and thus might provide examples of AGN in
the soft state with $\alpha_{\rm OX} \sim 2$.

We do not show the radio-loud
sample from Fig.~\ref{fig:fig1} here because of lack of the simultaneous
measurements of the photon index of the power-law like part of their X-ray
spectrum ({\it ROSAT} observations covered soft 0.1--2.4 keV X-ray band,
sufficient to estimate $\alpha_{\rm OX}$ but not $\Gamma$). However, we
characterize the radio-loud quasars in Fig.~\ref{fig:fig6} with average
properties derived for the composition of \citet[][ we plot the 2--10 keV
photon index]{elvisea:1994} and the sample of 16 RLQs of \citet[][
  $\alpha_{\rm OX}$  defined between 2500\AA\/ and 1 keV]{gambillea:2003}. On average the
RLQs are placed on the left low-$\Gamma$ edge of the GBH trail.
The derived average photon index of RLQs samples ranges from $\sim1.5$ to 1.8 and is flatter than average radio-quiet
$\Gamma = 2.06$ in \citet[][]{kelly:2007}. This effect is usually explained
with jet or jet-related emission. We can conclude that the RLQs would
also correspond to the intermediate/very high state of GBHs. We also show 13 low-luminosity LINERs from \citet[][]{maoz:2007}. We may hypothesize
that some radio-loud quasars \citep[e.g.][]{bechtold:94,gambillea:2003} and LINERs \citep[e.g.][]{maoz:2007} with $\alpha_{\rm OX} \sim 1$ and a typical
photon index of $\Gamma = 1.5$--2 \citep[e.g.][]{bechtold:94,belsole:06}
may be in a hard state.


\subsection{Dependence of spectral indices on disc and Comptonization fluxes}
\label{sec:relations}


\begin{figure*}
\centering
\includegraphics[height=6.0cm,bb=45 150 230 323,clip]{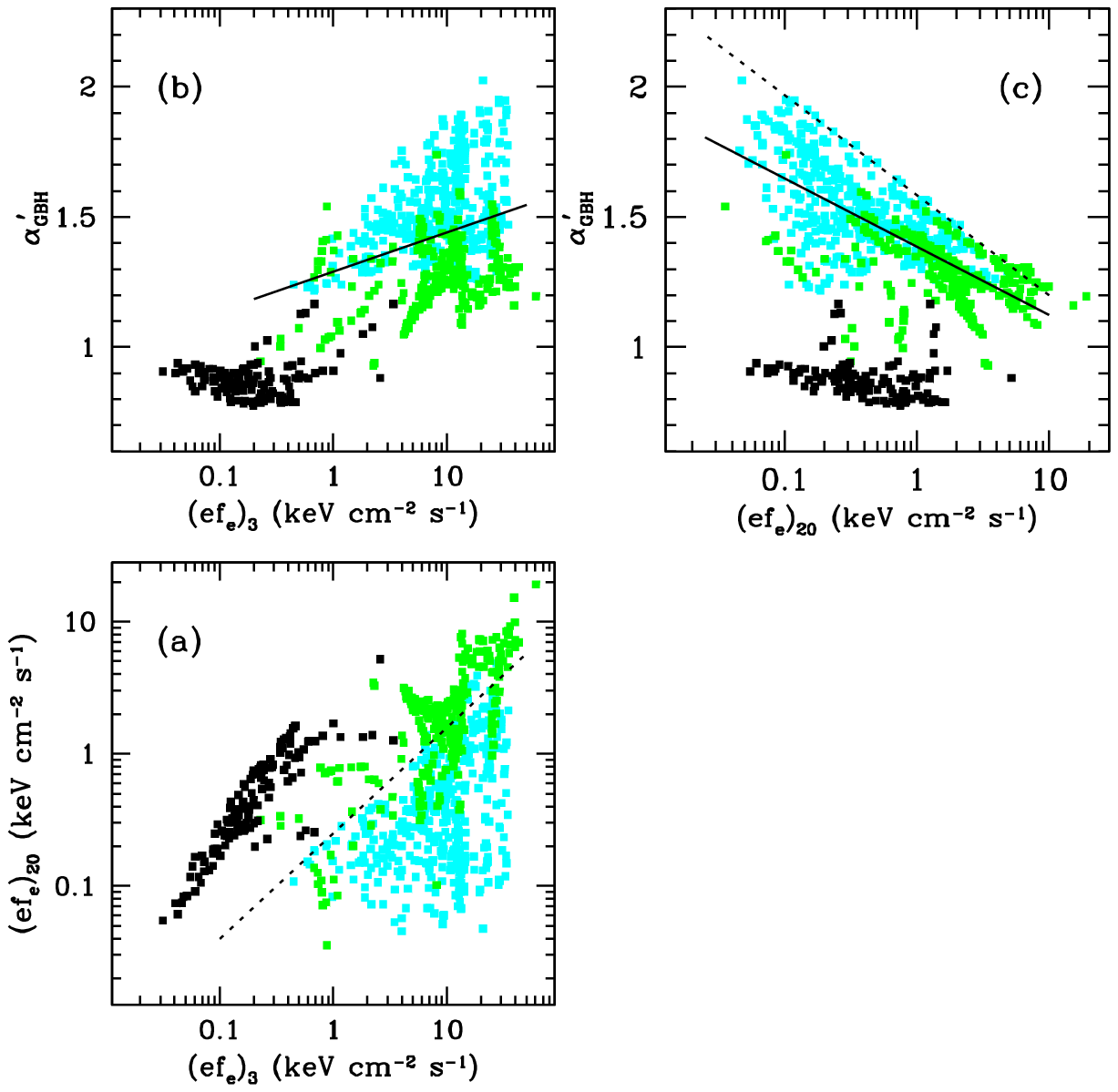}
\includegraphics[height=6.0cm,bb=45 320 230 500,clip]{fig7.ps}
\includegraphics[height=6.0cm,bb=260 320 405 500,clip]{fig7.ps}
\caption{Relations between the disc-to-Comptonization index, 
$\alpha_{\rm GBH}^{ \prime}$, and energy fluxes, ($ef_e$)$_{\rm 3}$,
  ($ef_e$)$_{\rm 20}$, for all GBHs considered in this 
paper based on Method 2 (hard state -- black, intermediate state --
dark grey/green, soft state -- light grey/cyan). (a) Comptonization flux 
vs. disc flux. The upper and lower
straight lines provide upper limits for the hard and soft state
data, respectively. (b, c) $\alpha_{\rm GBH}^{\prime}$ vs. disc and
Comptonization flux, respectively. The solid lines show formal correlations
between the quantities. The dotted line in (c) shows the upper limit on the
soft/intermediate state data.}
\label{fig:fig7}
\end{figure*}


Correlations between spectral indices ($\alpha_{\rm OX}$ and $\Gamma$) and
disc and Comptonization fluxes, $l_{\rm UV}$ and $l_{\rm X}$, are
intensively studied in the AGN samples. Particular attention is given
recently to the radio-quiet objects and here we will focus on this sub-class. Highly significant
correlation was reported between the optical/UV and 
X-ray fluxes \citep[e.g.][]{steffen:2006}, and between the $\alpha_{\rm
  OX}$ and $l_{\rm
  UV}$\citep[e.g.][]{steffen:2006,kelly:2007,kelly:2008}. \citet[][]{kelly:2008}
considered a large sample of objects 
with estimated mass of the black hole and they found that the X-ray photon
index $\Gamma$ correlates with both X-ray and optical/UV luminosities
expressed in Eddington units, i.e. $L({\rm 2 keV})/L_{\rm  Edd}$ and
$L({\rm 2500\AA})/L_{\rm Edd}$. They interpreted this as a result of $\Gamma$
correlation with the mass accretion rate.

In the previous sections we showed that there is no linear correlation of
$\alpha^{\prime}_{\rm GBH}$ with disc luminosity, $(ef_e)_{\rm 3}$, when
individual objects are considered. Here we check if there are any
correlations of $\alpha^{\prime}_{\rm GBH}$ with either disc or
Comptonization luminosities within a given spectral state, and for this
purpose we analyse data of all considered GBHs together. 

Figure~\ref{fig:fig7}a shows that in the hard state the $EF_E$ fluxes
at 3 and 20 keV are proportional to each other and that this
correlation may hold also for the intermediate state data.
This correlation arises because in the hard state the Comptonization
dominates the entire X-ray band (Fig.\ref{fig:fig2}a), including soft
X-rays which in other states are dominated by disc thermal emission, and so
$\log(ef_e)_{\rm 3}$ and $\log(ef_e)_{\rm 20}$ change in a similar
way. Such hard state correlation between the disc and Comptonization
fluxes should not be however expected in hard-state AGN whose disc
emission may be masked by the host galaxy.

The dotted line in Fig.~\ref{fig:fig7}a, $\log(ef_e)_{\rm 20} =
0.8\log(ef_e)_{\rm 3} - 0.6$, indicates the upper limit for
the soft state data. This limit shows a relation between the disc and
Comptonized components in the soft state. This state is dominated by
the accretion disc emission, so there is a threshold to the amount of
the Comptonization at a given thermal disc emission, e.g. a maximum to
the division of the energy release between the disc and the corona in
the accretion process.

The $\alpha^{\prime}_{\rm GBH}$ is consistent with being constant in the
hard state (Fig.~\ref{fig:fig7}b--c). In the
intermediate and soft states we find formal correlations of
$\alpha^{\prime}_{\rm GBH}$ with the disc and Comptonization fluxes,
$\alpha^{\prime}_{\rm GBH} = (0.150\pm0.018)\log{(ef_e)_{\rm 3}}+C_{\rm 1}$
and $\alpha^{\prime}_{\rm GBH} = (-0.262\pm0.010)\log{(ef_e)_{\rm
    20}}+C_{\rm 2}$, respectively. The former correlation is consistent with
those found for AGN \citep[][]{steffen:2006,kelly:2008}. The later is
marginally consistent with only the correlation reported by
\citet[][they consider, however, dependence of $\alpha_{\rm OX}$ on
monochromatic X-ray luminosity in Eddington units, and here we do not
correct GBH luminosities for difference in black hole mass]{kelly:2008}.

We do not see any obvious correlation between the photon index and either
the disc or Comptonization flux in any of the spectral states. This result
does not change if the fluxes are expressed in the Eddington units, as in
\cite[][]{kelly:2008}.

The intermediate state data may form narrow tracks due to
transitions between the hard and soft states (e.g. Figs.~\ref{fig:fig7}b-c).
The soft and intermediate states in Fig.~\ref{fig:fig7}c seem to be
restricted by the $\alpha^{\prime}_{\rm GBH} = -0.4\log(ef_e)_{\rm
  20}+C_{\rm 3}$
function. This limit is formed by the data with highest $(ef_e)_3$ flux
that for our sources does not exceed $\sim$40 keV cm$^{-2}$ s$^{-1}$ 
(Figs.~\ref{fig:fig7}a-b). Interestingly, the slope of this limit is
consistent with the $\alpha_{\rm OX}$ and $l_{\rm UV}$ correlation found by
\citet[][]{kelly:2008}. 

\section{Discussion and Conclusions}
\label{sec:conclusions}

In this paper we drew analogy between accretion disc vs. Comptonized
component emissions in GBHs and AGN. We introduced an $\alpha^{\prime}_{\rm
  GBH}$ parameter to describe the SEDs of GBHs throughout their
outbursts. This new parameter has the same physical meaning and occupies
the same range of values as the X-ray loudness, $\alpha_{\rm OX}$, used to
characterize the AGN optical-to-X-ray spectra. The central engine in both
AGN and GBHs is powered by a black hole accreting matter from the host
galaxy (AGN) or from a companion star (GBHs). Thus, the underlying physics
and radiative processes producing SED in these
two classes of objects should be similar. In particular, the properties of
$\alpha_{\rm OX}$ and $\alpha^{\prime}_{\rm GBH}$ distributions, and the
way they depend on other spectral observables should correspond to each
other.  

We have applied the methodology commonly used in the AGN studies to a
number of {\it RXTE} data of GBHs. We found that the majority of the
observed Type 1 radio quiet AGN may be in a spectral state
similar to a very high state, which is one of the soft states of
GBHs, or in an intermediate state. We base this conclusion on two
findings. First, detailed analysis of GBH outbursts
during which sources cover all spectral states shows that $\alpha^{
  \prime}_{\rm GBH}$ distribution has three distinctive peaks: $\sim$ 1
(corresponding to the hard spectral state), $\sim 1.5$ (very
high/intermediate state) and $\sim 2$ (typical soft and ultra-soft
state), while the $\alpha_{\rm OX}$ distribution for AGN is
single-peaked with an average close to 1.5
(Figs.~\ref{fig:fig5}~and~\ref{fig:fig1}). Second, the analysis
of the $\alpha^{ \prime}_{\rm GBH}$ vs. X-ray photon index,$\Gamma$,
diagram (Fig. \ref{fig:fig5}) reviled that Type 1 RQQs occupy the same region
as the GBHs in the 
intermediate/very high state. Interestingly, \citet[][]{mchardy:2007} reported that Ark 564 stays most probably in the very high/intermediate state. This conclusion was reached based on the variability studies and position of characteristic frequency in power density spectrum of this Seyfert 1. However, we note that a fraction of AGN from the \citet[][]{kelly:2008} sample reminds the soft state GBHs.

We argue that the observed radio-loud quasars may also be the
counter-parts of the very high/intermediate state GBHs. This is in
agreement with the GBHs studies that show presence of the radio emission in
some of the very high/intermediate states, shortly after the transition
from the hard (or typical soft) state
\citep[e.g][]{corbelea:2001}. We should note here that the peak in 
the distribution of $\alpha^{\prime}_{\rm GBH}$ around 1.5 is a
superposition of the radio-quiet and radio-loud states. In the
\citet[][]{fenderea:2004} jet model for GBHs the radio-quiet (soft)
and radio-loud (hard, intermediate, very high) states are separated with so
called 'jet line' placed at a certain value of hardness ration. In our
description, this would correspond to $\alpha^{\prime}_{GBH} \sim 1.4$--1.5
(the approximate border line between the intermediate/very high and soft state
data, Fig.~\ref{fig:fig7}b), which interestingly fits between the maxima
of the distributions of the $\alpha_{\rm OX}$ for the radio-loud and
radio-quiet AGN (Fig.~\ref{fig:fig1}).
More detailed discussion of the connection between the
radio-loud AGN and GBHs is however beyond the scope of the present paper. 
We note that spectra of LINERs show spectral indices similar to those seen in the hard state GBHs.

The presence of three peaks in the distribution of $\alpha^{\prime}_{\rm
  GBH}$ (Fig.~\ref{fig:fig5}) indicates that the time of transition between
  the spectral states is much shorter than the time the source spends in a 
given state. Thus, physical conditions resulting in characteristic
spectral shapes must be relatively `stable'. Analogous $\alpha_{\rm OX}$
distribution should be expected in AGN, and the most important question
arising from this study is whether AGN able to form the two peaks at
  $\alpha_{\rm OX}$ around 1 and 2 exist, but are missed in present surveys.

It is possible that the two extreme peaks apparent in the distribution
of $\alpha^{ \prime}_{\rm GBH}$ are missed in AGN samples
due to selection effects. The AGN samples rely on optical detections,
hence low disc luminosity hard state objects with $\alpha_{\rm
OX}\sim{1}$ could be absent from the sample. The AGN
optical flux in the hard state would be low and buried within the host
galaxy emission, and thus they could not be classified as AGN. However,
hard state AGN still should be present and 
even detected in X-ray surveys. It is interesting to note that they
may `mimic' so-called absorbed sources, as their X-ray spectra will
be hard. Their emission would contribute to the hard X-ray background
emission. The AGN counterparts of the ultra-soft GBHs with $\alpha_{\rm
  OX}\sim 2$ might have been omitted because their X-ray emission is
probably too weak to be detected by {\it Chandra} and {\it
  XMM-Newton}. This aspect is considered in more details in our next paper
(Sobolewska et~al., in preparation). 

All GBHs contain a black hole of similar mass, of the order of several
Solar masses \citep[see e.g.][ and references
therein]{donegier:03,mccremi:03}. In contrast, the AGN samples contain
black holes that differ in mass even by 3--4 orders of magnitude. It
is known that the mass of the central objects affects the temperature
of the accretion disc and thus alters the overall broad band energy
distribution. Thus, in Paper II we
discuss how the mass spread in AGN samples can affect the
results. In particular we check if the correlations between the
spectral indices and disc and/or Comptonization fluxes (not observed here,
or not consistent with the AGN observations) may arise as the effect of
mass distribution in the AGN samples. On the other hand, our results
presented in Sec.~\ref{sec:relations} provide predictions for relations
between the $\alpha_{\rm OX}$ and monochromatic fluxes in a narrow black
hole mass bin. Such studies may soon become possible as the number of AGN
for which black hole masses were estimated from independent methods
(gas/star dynamics, line widths, variability properties) increases.


\section*{Acknowledgements}
We thank Paul Green for valuable discussions about AGN surveys, and the
anonymous referee for careful reading of the manuscript and comments on how
to improve it. This research was funded in part by NASA contract
NAS8-39073 and Chandra awards GO2-3148A and GO5-6113X. MS was supported by
EU grant MTKD-CT-2006-039965 and by Polish grant
N20301132/1518 from Ministry of Science and Higher Education. MG acknowledges support from STFC Fellowship and 
Polish grant NN203065933.


\label{lastpage}

\end{document}